\documentclass[useAMS,usenatbib]{mn2e}

\usepackage{lipsum}
\usepackage{multicol}
\usepackage{graphicx}
\usepackage{multirow}
\usepackage{txfonts}

\usepackage{natbib}
\title[GX 339-4: spectral study with TCAF model]
{Characterization of GX 339-4 outburst of 2010-11: analysis by xspec using two component advective flow model}
\author[D. Debnath, S. Mondal and S. K. Chakrabarti]
{Dipak Debnath$^1$\thanks{E-mail: dipak@csp.res.in}, Santanu Mondal$^1$, Sandip K. Chakrabarti$^{1,2}$ \\
$^1$ Indian Centre For Space Physics, 43 Chalantika, Garia Station Road, Kolkata, 700084, India\\
$^2$ S. N. Bose National Center for Basic Sciences, JD-Block, Salt Lake, Kolkata, 700098, India}

\begin{document}

\date{Accepted 2014 December 4; Received 2014 October 27; MNRAS (February 21, 2015) 447 (2): 1984-1995.}

\maketitle

\begin{abstract}

We study spectral properties of GX 339-4 during its 2010-11 outburst with Two Component Advective
Flow (TCAF) model after its inclusion in XSPEC as a table model. We compare results fitted by TCAF
model with combined disk black body and power-law model. For a spectral fit, we use  $2.5-25$~keV 
spectral data of the PCA instrument onboard RXTE satellite. From our fit, accretion flow parameters
such as Keplerian (disk) rate, sub-Keplerian (halo) rate, location and strength of shock are extracted. 
We quantify how the disk and the halo rates vary during the entire outburst. We study how the halo 
to disk accretion rate ratio (ARR), quasi-periodic oscillations (QPOs), shock locations and its 
strength vary when the system  passes through {\it hard, hard-intermediate, soft-intermediate}, 
and {\it soft} states. We find pieces of evidence of monotonically increasing and decreasing 
nature of QPO frequencies depending on the variation of ARR during rising and declining phases.
Interestingly, on days of transition from hard state to hard-intermediate spectral state (during 
the rising phase) or vice-versa (during decline phase), ARR is observed to be locally maximum. 
Non-constancy of ARR while obtaining reasonable fits points to the presence of two independent 
components in the flow. 

\end{abstract}


\section{Introduction}

Galactic transient black hole candidates (BHCs) are very interesting objects to study in X-rays
because they exhibit rapid evolutions in their temporal and spectral properties during outbursts
\citep[see for reviews,][]{RM06,MR06}. 
In the past two decades, especially after the launch of {\it Rossi X-ray Timing Explorer} (RXTE), our 
understanding of black hole binaries has improved significantly. However, real progress in
extracting physical parameters was hampered due to lack of appropriate data analysis software. For instance, 
fitting a spectrum with a black body and a power-law component (so-called diskbb plus power-law, or 
compST models in XSPEC) tells us that there is a multicolor soft photon source such as a Shakura-Sunyaev 
standard disk \citep[][hereafter SS73]{SS73} 
and a so-called Compton cloud \citep{ST80,ST85} 
which is a hot region of free electrons with certain optical depth and temperature. However, cause of formation 
of the standard disk, nature and origin of Compton cloud or 
a specific spectral state remained missing. There was no information about why and how optical depths and temperatures of the 
cloud  or the accretion rates of the disk vary. A turning point in theoretical solutions of 
viscous transonic accretion flows around black holes \citep{C90a,C90b,C96} 
came when it was shown that flows above a critical viscosity parameter 
$\alpha$ will become a Keplerian  disk, while those below will remain sub-Keplerian. 
This fact was used to construct a two component advective flow (TCAF) model 
(\citealt{CT95}, hereafter CT95; \citealt{ETC96,C97}, hereafter C97), 
which not only explains under what circumstances a standard Keplerian disk could form, but also explains 
state transitions and variation of flow parameters during 
outbursts of several black holes reasonably well \citep{D08,D13,DC10,N12}.
In a TCAF solution, low viscosity and low angular momentum matter piles up behind a centrifugal 
barrier forming an axisymmetric shock (C90a,b; Molteni et al., 1994; Ryu et al. 1997), 
which is by and large stable even under non-axisymmetric perturbations \citep{Okuda07}. 
This barrier is known as the `CENtrifugal pressure dominated BOundary Layer', or CENBOL and acts as the Compton cloud. 
Regions of higher viscosity forms a Keplerian disk which settles down to a standard Shakura-Sunyaev 
disk when the cooling is efficient \citep[see,][and references therein]{GC13,GGC14}. 
Soft photons from SS73 disk are inverse Comptonized by the post-shock region (CENBOL) to form hard photons 
and thus in the TCAF (CT95) solution a separate Compton cloud is not required.
Another novel aspect is that fitting with TCAF does not require explicit knowledge of viscosity parameter.
TCAF directly uses two accretion rates, one for high viscosity flow (for super-critical viscosity 
parameter $\alpha$),  namely, standard  SS73 like Keplerian component (disk), 
and the other for low viscosity flow (sub-critical $\alpha$), namely, sub-Keplerian 
(low angular momentum) component (halo). Location and strength of 
shocks which TCAF uses also depend on viscosity, but since we use them as fitting parameters, a prior 
knowledge of viscosity is not essential. Furthermore, low-frequency quasi-periodic frequencies 
(QPOs) are supposed to be due to oscillations of the CENBOL region from where jets are 
originated \citep{C99}. 
So a separate oscillating component is not required to explain QPOs. Non-thermal
electrons are produced by the shock which may produce long power-law tail 
\citep{CM06} 
obviating the need to inject non-thermal electrons externally \citep{Zdziarski01}. 
Thus, TCAF aims to resolve all the spectral and temporal properties along with disk-jet connections within 
the framework of a single solution. These advantages motivated us to fit spectral data from BHCs 
with TCAF \citep{DCM14,MDC14}, even when they may be fitted equally well with other available models 
(such as diskbb plus power-law) in XSPEC. Other models do not discuss origin of the 
Compton cloud or corona \citep{Haardt93,Zdziarski03} 
and do not unify timing properties (e.g., QPOs) with spectral properties or 
obtain outflow rates with spectral properties as TCAF does. 
Unlike a persistent source, where accretion rates could be stable for a long time, 
in an outburst source, rates have to be varying. One of the most natural ways 
to achieve this is by varying viscosity. It is possible that
an outburst may be triggered by enhancement of viscosity at the outer disk. \citet{MC10} 
suggested that enhanced viscosity redistributes part of halo (i.e., low angular momentum flow) into a SS73 
like Keplerian flow, keeping the total mass-flow rate roughly constant. Their conclusion was that the accretion 
rate of the disk must go up rapidly due to enhanced viscosity before coming down when
viscosity is reduced again. This conversion and formation of TCAF configuration has 
been demonstrated recently by numerical simulations \citet{GC13}. 
A sudden increase in viscosity at the outer edge would progressively shift inner edge of the Keplerian disk
towards the black hole which causes a enhancement of soft luminosity. Declining phase starts when the source 
of enhanced viscosity is removed and there is a resulting shortfall of Keplerian component. Thus fitting an 
outburst source data with TCAF having two evolving accretion rates, shock location (CENBOL boundary) and shock 
strength may give us insight into how the disk structure really evolves. As mentioned earlier, though 
enhanced viscosity and its temporal and spatial variations are believed to be the prime cause of the outburst,
no prior knowledge of viscosity is required since the fitting parameters carry those information. In future,
we would pursue to derive even more fundamental parameters, such as, specific energy and angular momentum
distribution inside the flow and then computation of viscosity would be possible.

Galactic transient black hole candidate (BHC) GX 339-4 was first observed in 
1973 \citep{Markert73} 
by $1-60$ keV MIT X-ray detector onboard OSO-7 satellite. This stellar-mass black-hole binary has a mass function of 
$M_{bh}~sin(i)$ = $5.8\pm0.5~M_\odot$ and low-mass companion of mass $m$ = $0.52~M_\odot$ \citep{Hynes03,Hynes04}. 
This binary system is located at a distance of $d~\geq~6$~kpc \citep{Hynes03,Hynes04} 
with R.A.=$17^h02^m49^s.56$ and Dec.=$-48^\circ46'59''.88$. This recurring transient source has undergone five 
X-ray outbursts \citep{Nowak99,Belloni05,N12} 
during the RXTE era (in the period from 1996 to 2011). Based on evolution of their spectral and timing properties 
\citep{MR06,Belloni05,RM06,D13} 
which are also found to be correlated with a characteristic temporal evolution, namely, hardness-intensity diagram (HID) 
\citep{Maccarone03,Homan05}, 
various spectral states are identified during the past outbursts of the same source.
In general, there are mainly four basic states - {\it hard, hard-intermediate, soft-intermediate}, and {\it soft} 
states are observed during an outburst \citep{Homan05,D13,Motta09,N12}. 
In the literature, one can find extensive discussions on the properties of these spectral states \citep{vdK04,Belloni05,RM06,D08,D13}.
Complex outburst profile of BHCs begins and ends in hard/low hard state, keeping soft and intermediate 
states in between. It has been pointed out \citep{N12,D13} 
that these four basic spectral states form a hysteresis loop during their outburst phases in the sequence of 
{\it hard $\rightarrow$ hard-intermediate $\rightarrow$ soft-intermediate $\rightarrow$ soft $\rightarrow$ 
soft-intermediate $\rightarrow$ hard-intermediate $\rightarrow$ hard}. According to TCAF, hard states are 
formed when soft photons are unable to cool the Compton cloud (comprising collectively of CENBOL, pre-shock halo 
and outflow) while reverse is true for soft states. Intermediate states are a proof that accretion rates 
are comparable and they occur when viscosity is either rising or declining \citep[see also,][]{DC10}.

GX~339-4 showed X-ray activity of $17$~mCrab (in $4-10$~keV), observed on 2010 January 03 with the MAXI/GSC 
onboard ISS \citep{Yamaoka10}. 
The source remained active in X-rays for the next $\sim 14$ months and during this period, 
the source was monitored with RXTE, starting from 2010 January 12 \citep{Tomsick10}. 
Temporal and spectral properties of the source was studied extensively during this outburst by several authors 
(\citealt{Motta11,Stiele11,Shidatsu11a,D10}, hereafter Paper I; \citealt{N12}, hereafter Paper II).
Several attempts were made to explore multi-wavelength properties of the source during this outburst 
\citep{Rahoui12,Buxton12,Dincer12,Cadolle12}. 
Temporal as well as spectral variabilities and radio jets are observed during this outburst 
\citep{Corbel13a,Corbel13b, Yan12}. 
In {\it Paper I}, a preliminary result of timing and spectral properties during initial rising phase 
was presented. Subsequently, in {\it Paper II}, detailed timing and spectral study during the entire outburst 
was presented. In both these papers, spectral properties were studied with a combination of conventional thermal 
(disk black body) and non-thermal (power-law) model components. In order to understand detailed accretion flow dynamics, 
we need to fit with a more physical model, such as TCAF, which would enable us to extract actual physical 
parameters of the accretion flow. 

In this {\it paper}, for the first time, we show results of implementation of TCAF model in HEASARC's 
spectral analysis package XSPEC. Our goal is to find how the flow parameters 
vary from the beginning to the end of the outburst of GX~339-4 during its recent 2010-11 episode.

Since the number of available RXTE/PCA archival data points is too high, we choose a total of 
$50$ observations spread over $419$ days of the entire outburst, starting from 2010 January 12 
to 2011 March 6. We also 
compare our fitted spectral results with that of the conventional combined disk black body (DBB) 
and power-law (PL) model presented in Paper II. 
From our current study, we obtain variations of the flow parameters for the same states,
namely, {\it hard, hard-intermediate, soft-intermediate}, and {\it soft} as described in Paper II.
Here, we mainly concentrate on the rising and the declining phases of the outburst, where state transitions occur.
From the TCAF model fitted $2.5-25$~keV PCA spectra, we find that the variations of DBB and PL fluxes 
during the outburst as observed by \citet{N12} 
are consistent with variations of Keplerian (disk) rate and sub-Keplerian (halo) rate respectively.
We find some intriguing properties of the variation of {\it accretion rate ratio}    
(ARR, i.e., ratio between sub-Keplerian halo and Keplerian disk rates) 
and frequencies of quasi-periodic oscillations (LFQPOs; if observed)
during the days of class transitions. These will be discussed below and require further investigations. 

The {\it paper} is organized in the following way: in the next Section, we present a briefly describe the 
TCAF solution and how a spectrum is generated. In \S 3, we discuss observational results and 
data analysis procedures using HEASoft software. Here, we also mention methods of generation of TCAF model 
{\it fits} file used for spectral fittings. In \S 4, we present spectral analysis results obtained from 
TCAF model fits of RXTE PCA data and compare results with that of the combined DBB and PL model fits. 
We note special behaviour of the extracted flow properties when the object is in different spectral 
states, with a particular emphasis during state transitions. Finally, in \S 5, we present a brief 
discussion of our results and make concluding remarks.

\section {Model Description and Governing Equations}

A brief description of Two Component Advective Flow (TCAF) solution has been given in the Introduction. 
We use Chakrabarti-Titarchuk (1995; CT95) model code as the basic program for generating v0.1 of TCAF model 
{\it fits} file to fit black hole spectra. In the TCAF solution, an SS73 like Keplerian disk (high viscosity component) 
at the equatorial plane is immersed inside a low angular momentum sub-Keplerian (low viscosity) halo. The flow is 
considered to be axisymmetric. Though accretion rates would be a function of radial distance in a rapidly evolving 
system such as an outburst \citep{DC10}, 
each observational result is fitted with an effective accretion rate. The Keplerian disk, truncated 
at the shock location, locally emits a flux of radiation same as that produced by an SS73 disk. Geometry around the 
black hole is described by \citet{PW80} 
pseudo-Newtonian potential $\Phi_{PN}=-\frac{1}{2(r-1)}$, where $r$ is the radial distance of the flow in units 
of Schwarzschild radius ($r_g=2GM_{BH}/c^2$, $c$ being the velocity of light and $M_{BH}$ being the mass of the 
black hole). As a black hole accretion is necessarily transonic \citep[e.g.,][hereafter C90a]{PB81,PW80,C90a},
for a large region of the parameter space, the flow must jump from supersonic to subsonic branch, forming a 
standing or oscillating shock (C89; C90a,b)
before entering into a black hole through the inner sonic point. CT95 considers the region between the shock 
to inner sonic point, namely, the post-shock region (CENBOL), to be the Compton cloud. Contribution of 
the pre-shock halo, or winds were separately discussed there. In the present fits file, we do not include effects 
of optically thin pre-shock halo and the outflowing matter. The CENBOL cools down through inverse Comptonization 
and produces observed hard photons. Thus very cold CENBOL does not produce hard photons or outflows at all.
\citep[][and references therein]{C99,SC11}. 
In CT95 spectral properties are studied using only strong shocks since the idea there was to show that 
TCAF could explain spectral state transitions, in principle. Spectrum was generated for a set of 
Keplerian (disk) rates, sub-Keplerian (halo) rates and shock locations. 
Unlike other models \citep[][and references therein]{Zdziarski03}, 
TCAF does not require to add an extra `reflection' component, since this component is in built into the procedure. 
Albedo at each disk radius is included in computing disk reflection. Spectrum is calculated from intercepted photons 
by CENBOL region. Fraction of absorbed hard X-rays by the pre-shock Keplerian disk is used to heat it up and emit 
blackbody radiation at a higher temperature. The procedure is iterated till a convergence in all parameters is achieved. 
As given in CT95, we computed the effect by taking into account the photoelectric absorption and recoil effects. 
We took care of contribution of the first scattering exactly and of the multiple scattering using Fokker-Planck 
diffusion approximation \citep{Sobolev75}. 

In an outburst, the shock is expected to have a time varying compression ratio $R$, since the post-shock region
is expected to cool down as accretion rate increases. This, in turn, is governed by enhancement of viscosity. 
Thus, we use $R$ as a parameter. 
$R$ determines post-shock density (radial optical depth) and post-shock temperature, which, 
together with an assumption of vertical equilibrium gives height $h (X_s)$ (in $r_g$) of the flow at the shock $X_s$ 
(in $r_g$), an important quantity which determines percentage of soft photons intercepted by the 
'Compton cloud'. Flux $F_{ss}$ emitted from the optically thick Keplerian disk is obtained from SS73,
$$
\it F_{ss}=6.15\times10^{8}\dot{M}r^{-3}\Im(\frac{M_{BH}}{M_{\odot}})^{-2} ~~ergs~cm^{-2}s^{-1}.
\eqno{(1)}
$$
Here, $\Im = (1-3/r)^{1/2}$. In the above equation, mass of the black hole $M_{BH}$ is 
measured in units of mass of the Sun ($M_\odot$), disk accretion rate $\dot{M}$ is 
in units of $gm~s^{-1}$. Although this equation was derived with a no-torque condition at 
the inner stable circular orbit (ISCO) at $r=3$, it remains the same in our case, even though 
our SS73 disk is effectively evaporated inside the CENBOL and mixes with the sub-Keplerian 
flow when CENBOL is hot enough (in hard states). In soft states, the CENBOL itself 
is cold and Keplerian disk can reach close to ISCO. \citet{Shidatsu11b}, calculated 
innermost disk radius for the black hole candidate GX~339-4,
which is consistent with the accepted picture of the high/soft state in which standard disk may
always extend up to ISCO. However, \citet{Allured13}
examined inner disk radius using different models (e.g., thermal
and relativistic line broadening etc.) at different luminosities in the hard states and found
that the disk need not go up to ISCO. In fact, since a black hole accretion is always 
supersonic, a Keplerian disk will be almost sonic/transonic at ISCO 
to prepare itself for the final plunge into the horizon with a radial velocity same 
as that of the velocity of light. So, disk extending up to ISCO is always an assumption, 
more so for low accretion rates. This is also evident from viscose transonic flow solutions (C90ab).
Angular momentum is transported inside the CENBOL till the inner sonic point at $r_c\sim 2.5 r_g$ 
\citep[see,][]{C90a}, 
where, ideally a torque-free condition should be applicable. Inside $r_c$, angular momentum transport 
is negligible. In future, we plan to actually simulate Keplerian disk along with 
the halo and compute time dependent spectra and several assumptions made here would 
become unnecessary. Already we have reported preliminary results of our 
simulations \citep{GC13,GGC14}. 
In the simulation, temperature would be obtained from the global solution of an advective flow,
and not as in SS73 where local gravitational energy dissipation was used. 

The post-shock region becomes hot due to conversion of kinetic energy of the pre-shock 
flow into thermal energy. However,      
electrons lose energy due to bremsstrahlung and Comptonization of soft photons emitted from the Keplerian disk.
Energy equation which protons and electrons obey in the post-shock region is given by (CT95),
$$
\frac{\partial(\varepsilon+\frac{P}{\rho})}{\partial r}+(\Gamma - \Lambda)=0,
\eqno{(2)}
$$
where, specific energy \citep{C89} 
of the flow is given by, 
$$
\varepsilon = \frac{v^2}{2} + na^{2} + \frac{\lambda^{2}}{2r^{2}} - \frac{1}{2(r-1)},
\eqno{(3)}
$$ 
and $\Gamma$ and $\Lambda$ are the heating and the cooling terms, respectively. Here, $\lambda$ is the specific 
angular momentum in units of $c r_g $, $v$ is infall velocity in units of $c$, $n=\frac{1}{(\gamma-1)}$ 
is polytropic index, $a = \surd(\frac{\gamma P}{\rho})$ is the sound speed of the medium in units of $c$, 
and $\gamma$ is adiabatic index. Equation (2) was not used in pre-shock halo region. 
Synchrotron radiation was not used in the present version of our fits file.

Since temperature and height of the post-shock region depend on the shock location and strength,
it is more physical to write them as functions of flow parameters. For this, we use pressure balance 
condition, where sum of the thermal pressure and the ram pressure must match on both sides of the 
shock \citep{Landau59}. 
In pressure balance condition (C89), we assume that the pre-shock flow is cooler, $a \sim 0$ as it falls freely and 
$X_s^2 \gg \lambda^2$. Using adiabatic $a$, we get height of shock front as, 
$h_{shk}= a~X_s^{0.5}~(X_s-1)$ 
for vertical equilibrium model. Initial guess of the shock temperature is obtained from the definition 
$a_s^2=\frac{\gamma kT_{shk}}{\mu m_p}$. Thus, temperature ($T_{shk}$) and height ($h_{shk}$) of the shock becomes:
$$
T_{shk} = \frac{m_p (R-1)c^{2}}{2R^{2}k_{B}(X_{s}-1)},
\eqno{(4)}
$$ 
$$
h_{shk} = \surd(\frac{\gamma (R-1) {X_{s}}^{2}}{R^{2}}), 
\eqno{(5)}
$$ 
where, $m_p$, $R$, $k_{B}$ and $X_{s}$ are mass of the protons, compression ratio, Boltzmann constant 
and shock location of the flow respectively. Soft photons from 
the pre-shock Keplerian disk are intercepted by the hot CENBOL. As the accretion rate increases, number of 
scattering increases and photons gain more energy while cooling down the CENBOL. 
Details about the solution procedure and enhancement factor calculation due to Comptonization 
is in Eq. (12a-14) of CT95. The outcome depends on the opacity, 
$$
\tau_T = \int_{r_i}^{X_s} \sigma_T n_e dr.
\eqno{(6)}
$$
Here, $r_i$ is the inner radius of CENBOL $\sim 2.5r_g$. Average energy exchange per scattering is given by 
($h\nu, kT_e \ll m_e c^2$),
$$
\frac{\Delta \nu}{\nu} = \frac{4kT_e - h\nu}{m_e c^2}.
\eqno{(7)}
$$
When $h\nu \ll kT_e$, photons gain thermal energy. Here, terms $n_e$, $\sigma_T$, $T_e$ and $r_i$ used in above 
two equations are number density of electrons in the post-shock region, Thomson scattering cross-section, average 
temperature of electrons and inner edge of the flow respectively. Balance of scattering and energy gain of photons 
give the power-law distribution as,
$$
F_{\nu} \propto \nu^{-\alpha}. 
\eqno{(8)}
$$

As mentioned already, to generate the model spectra (which are used as inputs for generating
TCAF model {\it fits} file), we made several modifications in original CT95 code to include,
\begin{enumerate}
\item [$i)$] Variation of compression ratio $R$ is allowed from $4$ (strong) to $1$ (weak). 
CT95 assumed only strong shock for illustration purpose.
\item [$ii)$] Computation of temperature of post-shock region using this $R$.
\item [$iii)$] Radial velocity of a rotating flow as in C97. 
\item [$iv)$] Spectral hardening correction of \citet{ST95}, 
which depends on the accretion flow rate. We uniformly consider the correction factor ({\it f}) to be
$1.8$ to calculate effective temperature in emitted spectrum. 
\end{enumerate}

We wish to mention certain limitations of the present version: although by taking extreme limits of 
Keplerian rate ($\dot{m_d}$ in Eddington rate) or sub-Keplerian rate ($\dot{m_h}$ in Eddington rate)
as close to zero as possible, we could have extreme hard or soft states, 
these limits require finer model grids. In the present version (v0.1), we do not consider cases with 
$\dot{m_h} \rightarrow 0$ (soft states). However, cases with $\dot{m_d} \rightarrow 0$ are not needed 
as spectral index is highly insensitive to ${\dot{m_d}}$ near this limit \citep{ETC96}. 
We also do not consider bulk motion Comptonization \citep[see, CT95;][]{Titarchuk98}. 
Present model of TCAF does not include magnetic fields explicitly or production of non-thermal 
photons by the shock. Thus inverse Comptonization of non-thermal photons produced in 
the post-shock region could not be included and thus some features 
which may occur at much higher energies in different sources 
\citep{Zdziarski01,CM06} 
cannot be fitted with the current {\it fits} file. All these points will be taken up in future.

\section {Observation and Data Analysis}

We present spectral analysis results of publicly available archival data from RXTE Proportional 
Counter Array (PCA) instrument for entire 2010-11 outburst of GX 339-4, starting from $2010$ January $12$ (Modified 
Julian date, MJD=$55208$) to $2011$ March $6$ (MJD=$55626$) from the PCA (Jahoda et al., 1996). 
In general, we follow the same analysis techniques as discussed in Paper II, for extraction of source 
and background `.pha' files using {\it Standard2} mode science data of PCA (FS4a*.gz). $2.5-25$ keV 
PCA background subtracted spectra are fitted with TCAF model {\it fits} file in XSPEC v. 12.8. 
To achieve the best fit, a Gaussian line of peak energy around $6.5$~keV (iron-line emission) is used. 
For the entire outburst, we keep hydrogen column density (N$_{H}$) for absorption model {\it wabs} fixed 
at 5$\times$~10$^{21}$~atoms~cm$^{-2}$ \citep{Motta09} 
and assume a $1.0$\% systematic error. After achieving the best fit based on reduced chi-square ($\chi^2_{red}$) 
value ($\leq 2$), to find 90\% confidence error values for TCAF model fitted parameters, `err' command is used 
(except data for rising soft-intermediate and soft states, from MJD = 55316 to MJD = 55593, $1~\sigma$ error 
are given, since here reduced $\chi^2$ values are found to be in between $1.8-2.8$). In Appendix I, detailed 
spectral fitted analysis results with observed QPO frequencies are provided (Note that error values for TCAF 
model fitted parameters which are given in Appendix I, are average values of 90\% confidence $\pm$ error, 
or, $1~\sigma$ error). 

One can also fit a spectrum by manually comparing observational spectrum with the theoretical model spectra, 
generated by different input parameters in TCAF model source code \citep{DC10}. 
However, in order to fit spectra more accurately one needs to use it in a complete package like XSPEC, which 
automatically achieves the best fit by iterative least square fit technique. From the spectral fit, one can obtain 
model fitted values of reduced chi-square, degrees of freedom, parameter errors etc. In order to 
fit black hole spectra with the TCAF model in XSPEC, we have generated model {\it fits} file 
({\it TCAF0.1.fits}) using theoretical spectra generated by varying five input parameters in CT95 code 
(after modifications mentioned earlier) and included it in XSPEC as an additive table model. These parameters 
are: $i)$ black hole mass ($M_{BH}$) in solar mass ($M_\odot$) unit, 
$ii)$ Keplerian rate ($\dot{m_d}$ in Eddington rate, $\dot{M}_{Edd}$), $iii)$ sub-Keplerian rate 
($\dot{m_h}$ in $\dot{M}_{Edd}$), 
$iv)$ location of shock ($X_s$ in Schwarzschild radius $r_g$), and $v)$ compression ratio 
($R$) of the shock. Of course, the model normalization value ($norm$), which for simplicity could be written as
$\frac{R_z^2}{4\pi D^2} sin(i)$, where, `$R_z$' is the effective height of the Keplerian component 
in $km$ at the pre-shock region,
`$D$' is source distance in $10$~kpc unit and `$i$' is disk inclination angle with the line of sight is also a variable. 
However, since $R_z$ need not come out of SS73 model due to its proximity of the shock surface, we
leave it as a free parameter to be determined from the best fitted value, very much like $R_{in}$, the inner 
edge of the disk, as obtained from the normalization of the diskbb model fits. In future, when we have 
a clearer picture of how $R_z$ should be computed theoretically, we will create a fits file with this to be the sixth parameter.

\begin{table}
\vskip 0.0cm
\addtolength{\tabcolsep}{-5.0pt}
\small
\centering
\caption{\label{table1} TCAF model {\it fits} files generated with different sets of model input parameter grids}
\vskip 0.0cm
\begin{tabular}{|l|ccc|ccc|ccc|ccc|ccc|}
\hline
Set &\multicolumn{3}{|c|}{$M_{BH}$ ($M_\odot$)} &\multicolumn{3}{|c|}{$\dot{m_d}$ ($\dot{M}$$_{Edd}$)} & \multicolumn{3}{|c|}{$\dot{m_h}$ ($\dot{M}$$_{Edd}$)} & \multicolumn{3}{|c|}{$X_s$ ($r_g$)} & \multicolumn{3}{|c|}{R} \\
    & Min & Max & GN & Min & Max & GN & Min & Max & GN & Min & Max & GN & Min & Max & GN \\
\hline
I   & 3 & 15 & 6 & 0.1 & 12.1 & 18 & 0.1 & 12.1 & 18 & 6 & 456 & 20 & 1 & 4 & 10 \\ 
II  & 3 & 15 & 6 & 0.1 &  2.0 & 15 & 0.1 &  2.0 & 15 & 6 & 456 & 24 & 1 & 4 & 12 \\ 
III & 3 & 15 & 6 & 2.0 & 12.1 & 15 & 2.0 & 12.1 & 15 & 6 & 456 & 24 & 1 & 4 & 12 \\ 
IV  & 3 & 15 & 6 & 0.1 &  2.0 & 12 & 0.1 & 12.1 & 20 & 6 & 456 & 24 & 1 & 4 & 12 \\ 
V   & 3 & 15 & 6 & 0.1 & 12.1 & 20 & 0.1 &  2.0 & 12 & 6 & 456 & 24 & 1 & 4 & 12 \\ 
\hline
\end{tabular}
\leftline {Here Min, Max, and GN represent minimum, maximum and grid numbers}
\leftline {for logarithmic equi-spaced model input parameters. }
\label{table1}
\end{table}

Five model input parameters mentioned above were varied in the following ranges: $i)$ $M_{BH}$: 3 -15 $M_\odot$, $ii)$ $\dot{m_d}$: 0.1 - 12.1 $\dot{M}_{Edd}$, $iii)$ $\dot{m_h}$: 0.1 - 12.1 $\dot{M}_{Edd}$, $iv)$ $X_s$: 6 - 456 $r_g$, and 
$v)$ $R$: 1 - 4, respectively. We first generate $\sim 4\times10^5$ model spectra by varying input parameters in above 
mentioned limits and then these model spectra are used as input files to a program written in FORTRAN, to 
generate a crude grid based model {\it fits} file (Set I of Table 1). 
However, once a reasonable fit is obtained with this {\it fits} file, spectra are refitted with appropriate finer grid 
based {\it fits} file (Set II-V of Table 1), to have a better fit with minimum parameter error values as well as better 
reduced $\chi^2$. This two step process is presently needed as we do not have a very fine grid for the entire parameter 
space. When we build the model with final grids around $\dot{m_h} \rightarrow 0$ also, fitting of soft states would 
be addressed satisfactorily. In the fitted results below, we froze the mass to be $5.8$ and therefore only 
four parameters were varied.

\section {Results of Fitting of Data by TCAF Solution}

Detailed temporal and spectral properties of the source during its 2010-11 outburst is already presented by several authors 
\citep[][Paper I and Paper II]{Motta11,Stiele11,Shidatsu11a,Cadolle12}.

Presently, we fit with {\it TCAF0.1.fits} file as generated above and compare our spectral results with 
combined DBB and PL model fitted results presented in Paper II. Here we discuss how the flow parameters evolve 
during rising and declining phases of the outburst. Detailed analysis results are given in Table Appendix  I,
where we present Observations IDs (Col. 2), date of observation (Col. 3), accretion rates (Cols. 4-5), accretion rate
ratios (Col. 6), shock locations (Col. 7), Compression Ratios (Col. 8), shock heights (Col. 9), temperature
at the CENBOL surface from Eq.(4) (Col. 10), thickness $R_z$ of the pre-shock Keplerian disk in km (Col. 11), line width and  
line depths (Cols. 12-13) of the iron line (where applicable), QPO frequencies (Col. 14) 
and finally $\chi^2/$DOF (Col. 15).

\begin{figure}
\vskip -0.0cm
\centering{
\includegraphics[scale=0.6,angle=0,width=9.0truecm]{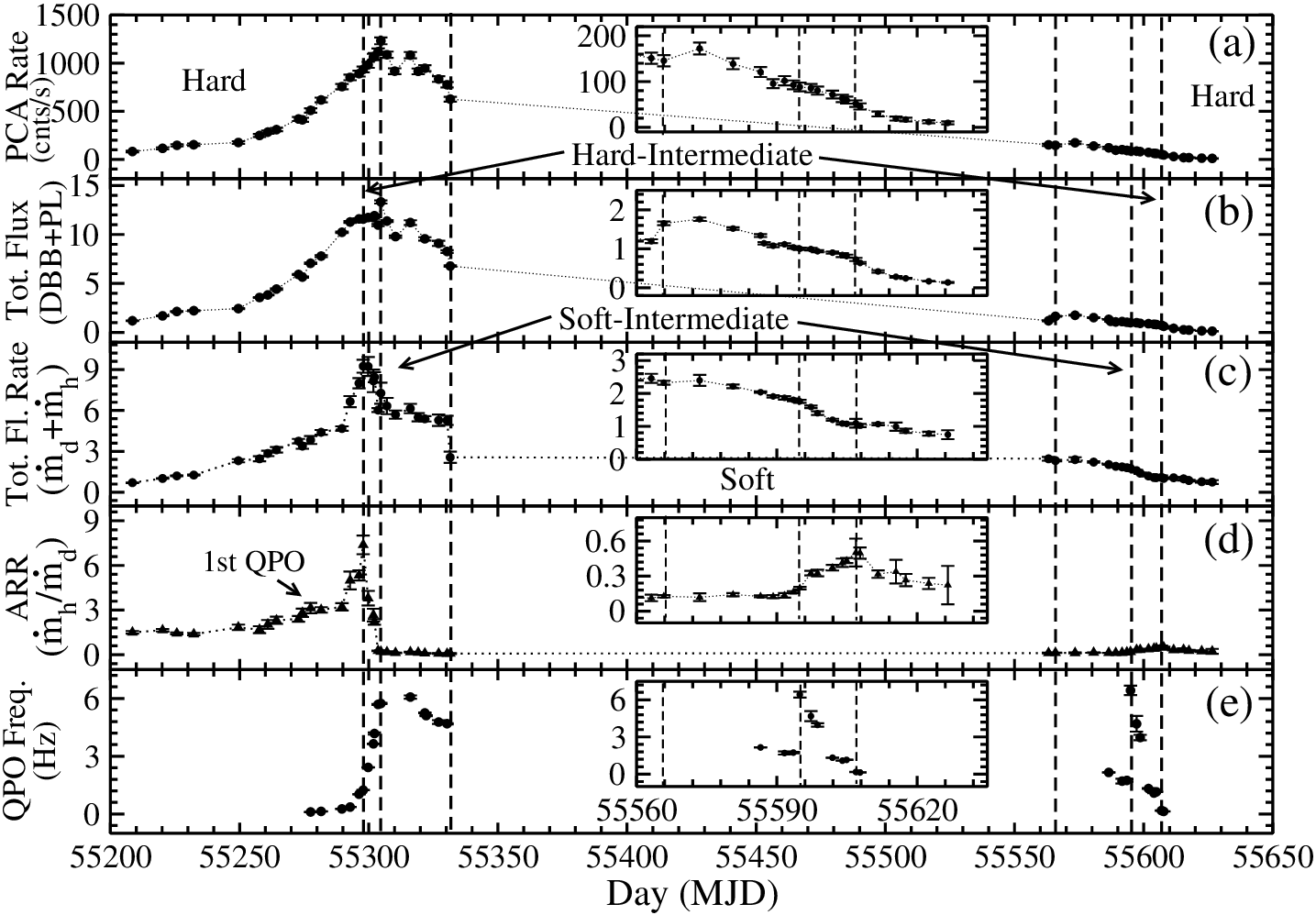}}
\caption{Variation of (a) $2-25$~keV PCA count rates (cnts/sec), (b) combined disk black body (DBB) and power-law 
(PL) model fitted total spectral flux in $2.5-25$~keV range (in units of $10^{-9}~ergs~cm^{-2}~s^{-1}$), 
(c) TCAF model fitted total flow (accretion) rate (in $\dot{M}$$_{Edd}$; sum of Keplerian disk, $\dot{m_d}$ and 
sub-Keplerian halo $\dot{m_h}$ rates) in the $2.5-25$~keV energy band, and (d) Accretion Rate Ratio (ARR; ratio 
between halo and disk rates) with day (MJD) for the 2010-11 outburst of GX~339-4 are shown. 
In the bottom panel (e), observed primary dominating QPO frequencies (in Hz) with day (MJD) are shown. 
The vertical dashed lines indicate the transitions of between different spectral states. Zoomed quantities in the
decay phase are in the inset.}
\label{fig1}
\end{figure}

\subsection{Evolution of fitted parameters during the outburst}

In Fig. 1a, variation of background subtracted RXTE PCA count rate in $2-25$~keV ($0-58$ channels) energy band 
with day (MJD) is shown. In Figs. 1b \& 1c, variations of combined DBB and PL model fitted total spectral flux 
(flux contributions for the DBB and PL model components are calculated by using the convolution model `cflux' 
technique after fitting a spectrum with combined model components) in $2.5-25$~keV energy band (Paper II) and 
TCAF model fitted total accretion rates (combined Keplerian disk and sub-Keplerian halo rates) in the same energy 
band with day (MJD) are shown. In Fig. 1d, variation of {\it Accretion Rate Ratio} ARR (defined as the
ratio of sub-Keplerian halo $\dot{m_h}$ and Keplerian disk $\dot{m_d}$ rates) with day (MJD) 
is shown. As will be evident from the plots of Fig. 3, it may be better to treat ARR as a proxy to 
hardness ratio since the latter is not a black hole mass independent concept. 
We also zoom a part of ARR plot in the declining phase to see their behaviour 
clearly and to emphasize that it reaches a maximum on the day when the hard state starts.
Observed QPO frequencies are shown in Fig. 1e. From the variation of ARR and nature (shape, frequency, $Q$ value, 
rms\% etc) of QPO (when observed) in the four different spectral states, namely, {\it hard, hard-intermediate, 
soft-intermediate, soft}, we find a distinct pattern. On or about the  days of spectral transitions between states 
as reported in Paper II (marked with vertical dashed lines in Figs. 1-3), ARR shows interesting behaviour (to be 
discussed in detail below). We also see that TCAF model holds good in fitting the spectra (model fitted 
$\chi^2_{red}$ varies $\sim 0.9-1.6$) from hard, hard-intermediate and soft-intermediate (declining) spectral 
states. In rising soft-intermediate and soft 
spectral states, TCAF model fitted $\chi^2_{red}$ varies in the range $\sim 1.8-2.8$. However, for a soft 
spectral state, where the source stayed for almost $\sim 8$ months, TCAF model cannot be used since we essentially 
need a single component with a high rate (outside of our grid used in this fits file).
In future, we will generate model {\it fits} file with a larger parameter range including $\dot{m_h} \rightarrow 0$, 
high $\dot{m_d} \geq 12$ and bulk motion Comptonization. 
In Fig. 2(a-e), variations of TCAF model fitted shock parameters, namely, $T_{shk}$ in $10^{10}$ K, and $X_s$ in units 
of $r_g$, $h_{shk}$ in units of $r_g$, the ratio $h_{shk}/X_s$ and $R$ are shown as a function of time (MJD).
During the initial rising phase of the outburst, as the day progresses, the shock generally moves 
towards the black hole (see, Fig. 2b) while progressively becoming weaker (see, Fig. 2e) due to increasing cooling. 
At the same time, shock height ($h_{shk}$) and temperature ($T_{shk}$) increase initially, then start to decrease gradually (Fig. 2 c-d) .
The ratio of shock height and shock location is around $0.7$ for a long 
interval of time when the object was in a hard state to about $0.3-0.4$ in the soft state. 
Thus the geometry of the CENBOL is sometimes spherical and sometimes disk like
as far as Comptonization process goes. So, we use Eq. (9) of \citet{Hua95} 
which is valid for two regimes to obtain spectral index. To obtain an average temperature of the CENBOL, 
we took weighted average of the temperature variation assuming it to be spherical. The processes
followed are exactly as in CT95. 

In the declining phase, almost an opposite nature of variations of these four parameters are observed. 
In Fig. 3(a-b), variations of combined DBB and PL model fitted DBB flux (top panel) and TCAF model fitted 
Keplerian disk rate $\dot{m_d}$ with day (MJD) are shown. Similarly, in Fig. 3(c-d), variations of combined 
DBB and PL model fitted PL flux and TCAF model fitted sub-Keplerian halo rate $\dot{m_h}$ (bottom panel) 
with day (MJD) are shown. Observed variations of these two different types of model fitted fluxes/rates are 
similar in nature.

Note that the pre-shock Keplerian disk thickness $R_z$ which comes out from the fit (Col. 11) is not necessarily 
decreasing with the reduction of the shock. This is because the increase in accretion rate caused radiation 
pressure to increase which puffs up the inner disk.

\begin{figure}
\vskip -0.1cm
\centering{
\includegraphics[scale=0.6,angle=0,width=9.0truecm]{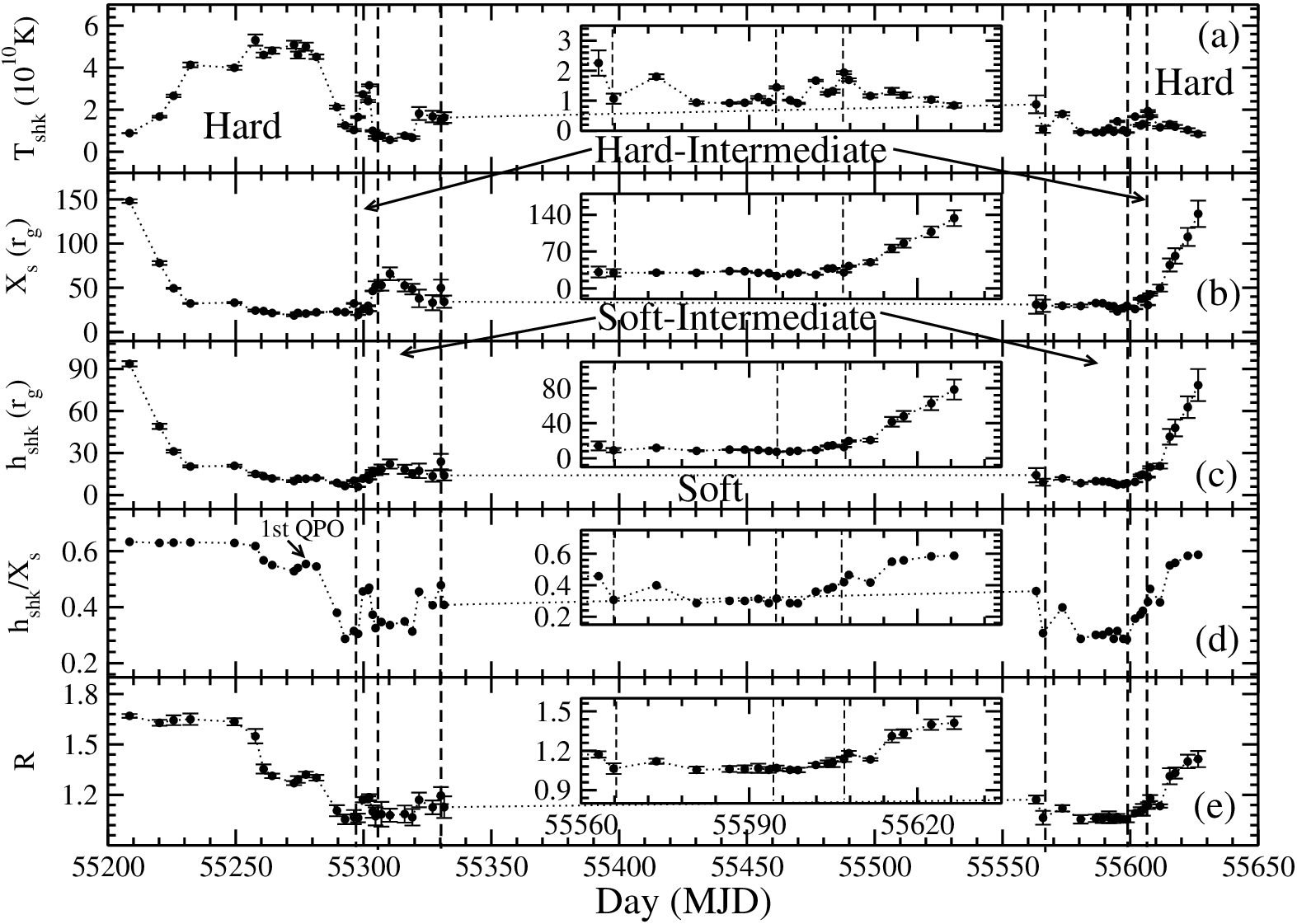}}
\caption{Variation of TCAF model fitted/derived shock (a) temperature ($T_{shk}$ in $10^{10}$ K), (b) location ($X_s$ in $r_g$), (c) height ($h_{shk}$ in $r_g$), (d) ratio between $h_{shk}$ \& $X_s$, 
and (e) compression ratio ($R$) with day (MJD) are shown. The shock height and temperature are 
calculated using Eqs. 4 \& 5 respectively.
}
\label{fig2}
\end{figure}

\subsection{Evolution of spectral and temporal properties during the outburst}

In Figs. 4(a-f), TCAF model fitted $2.5-25$~keV background subtracted PCA spectral flux, $E^2 f(E)$ 
in units of $keV^2~(Photons~cm^{-2}~s^{-1}~keV^{-1})$ with variation of $\Delta \chi$ in three different 
(hard, hard-intermediate, soft-intermediate) spectral states in both rising 
and declining phases are shown. These results are presented in Table 'Appendix I'.


Generally, low frequency quasi-periodic oscillations (LFQPOs) in $\sim 0.01-30$~Hz are observed during 
hard and intermediate spectral states of transient black hole candidates 
\citep[see,][for a review]{RM06}. 
From detailed temporal and spectral study (using a combined DBB and PL model fits) of present 2010-11 
outburst of GX~339-4 (Paper I and Paper II) and 2010 \& 2011 outbursts of H~1743-322 \citep{D13}, 
it has been observed that QPO frequency increases monotonically during rising hard and hard-intermediate 
spectral states, and decreases monotonically through the same spectral states in declining phases. 
It has also been observed that during soft-intermediate spectral states of both rising and declining 
phases, QPOs are observed sporadically, whereas no LFQPOs are observed in the soft spectral state. 
According to Paper II and \citet{D13}, 
maximum value of evolving QPO frequencies (fitted with Propagating Oscillatory Shock (POS) model solution) 
are observed on the final day of rising hard-intermediate spectral state {\it and} on the very first day 
of the same spectral state in declining phase of outburst. 
From our spectral study of present outburst of GX~339-4 using TCAF model, we notice that on these days when 
the observed QPO frequency is maximum, ARR reaches its lower values, where source stayed for a longer 
durations (in soft-intermediate or soft spectral states) with little change in the rising or releases from 
from its long stayed Keplerian disk dominated lower values to increase rapidly in the declining phases 
of the outburst (Fig. 1d-e). It also has been noticed that during the rising phase of the outburst, QPO 
frequency starts to increase rapidly when ARR reaches its peak value and during the declining phase QPO 
frequency starts to decrease slowly (or leaves its rapid decrease phase) when ARR reaches its maximum value 
(of the declining phase) of the outburst (Fig. 1d-e).

\begin{figure}
\vskip -0.1cm
\centering{
\includegraphics[scale=0.6,angle=0,width=9.0truecm]{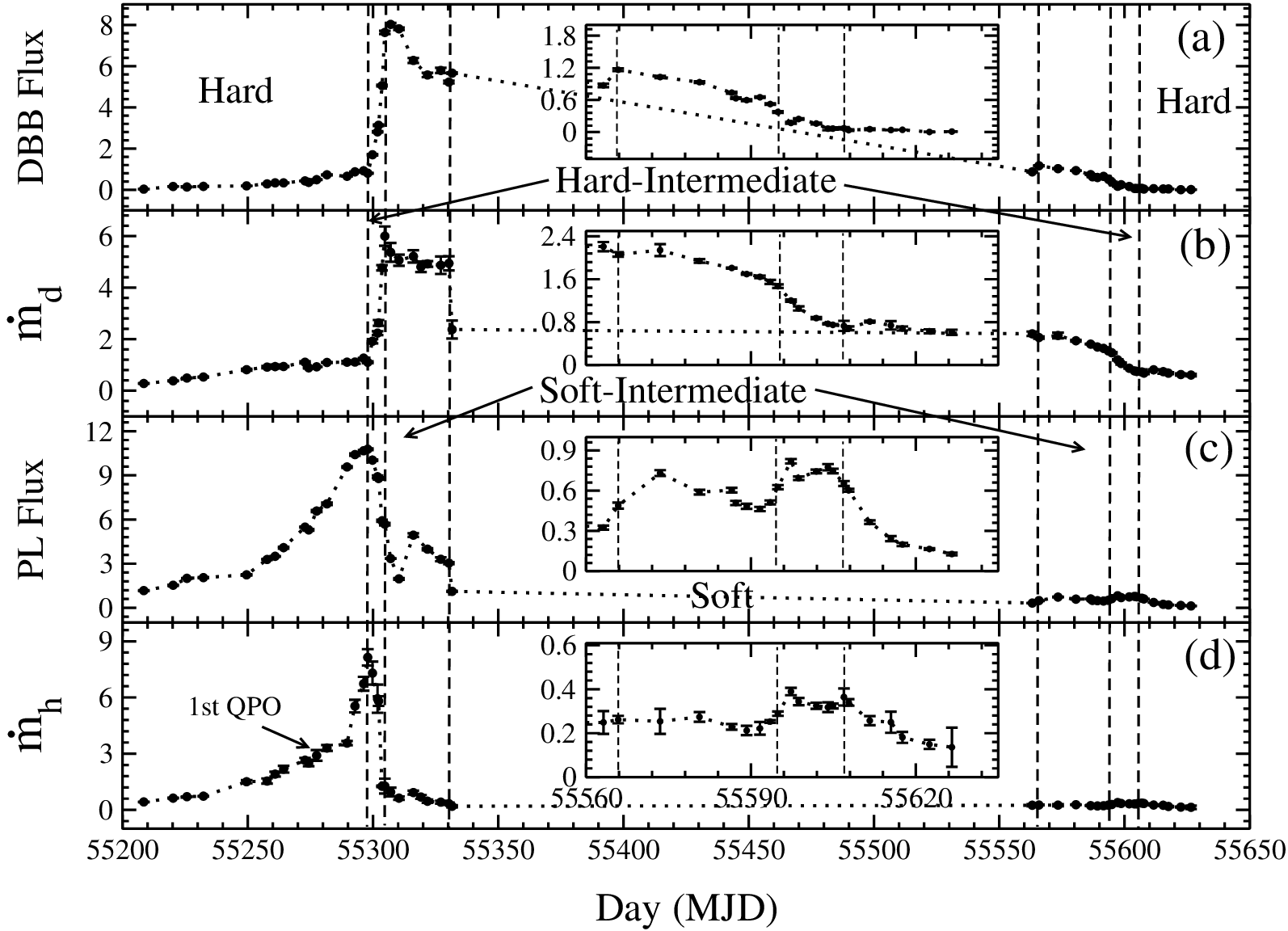}}
\caption{In top panel (a), the variation of combined disk black body (DBB) and power-law (PL) model fitted DBB 
spectral flux and in panel (c), the variation of PL spectral flux (both in units of $10^{-9}~ergs~cm^{-2}~s^{-1}$) 
in $2.5-25$~keV energy range are shown. In panel (b), the variation of TCAF model fitted Keplerian disk rate $\dot{m_d}$ 
and in bottom panel (d), the variation of sub-Keplerian halo rate $\dot{m_h}$ (both in $\dot{M}$$_{Edd}$) in the
same energy band are shown. 
}
\label{fig3}
\end{figure}

In Papers I \& II, spectral classifications were discussed based on the degree of importance of 
DBB and PL model components (according to variation of fitted component values and their individual fluxes) 
and nature (shape, frequency, $Q$ value, percentage of rms amplitude etc.) of QPO (if present). However, here we
find that ARR varies in a well defined way in these states. In the rising phase, ARR increases steadily in the hard state, 
decreases very fast in the hard intermediate state, remains almost constant in the soft-intermediate state. Exactly
opposite behaviour is in the declining phase. It is also clear that four basic spectral states 
form a hysteresis loop during the outburst, in that the way the accretion rates change 
in the rising phase is different from that in the declining phase.

\begin{figure}
\centerline{
\includegraphics[scale=0.6,angle=270,width=4.5truecm]{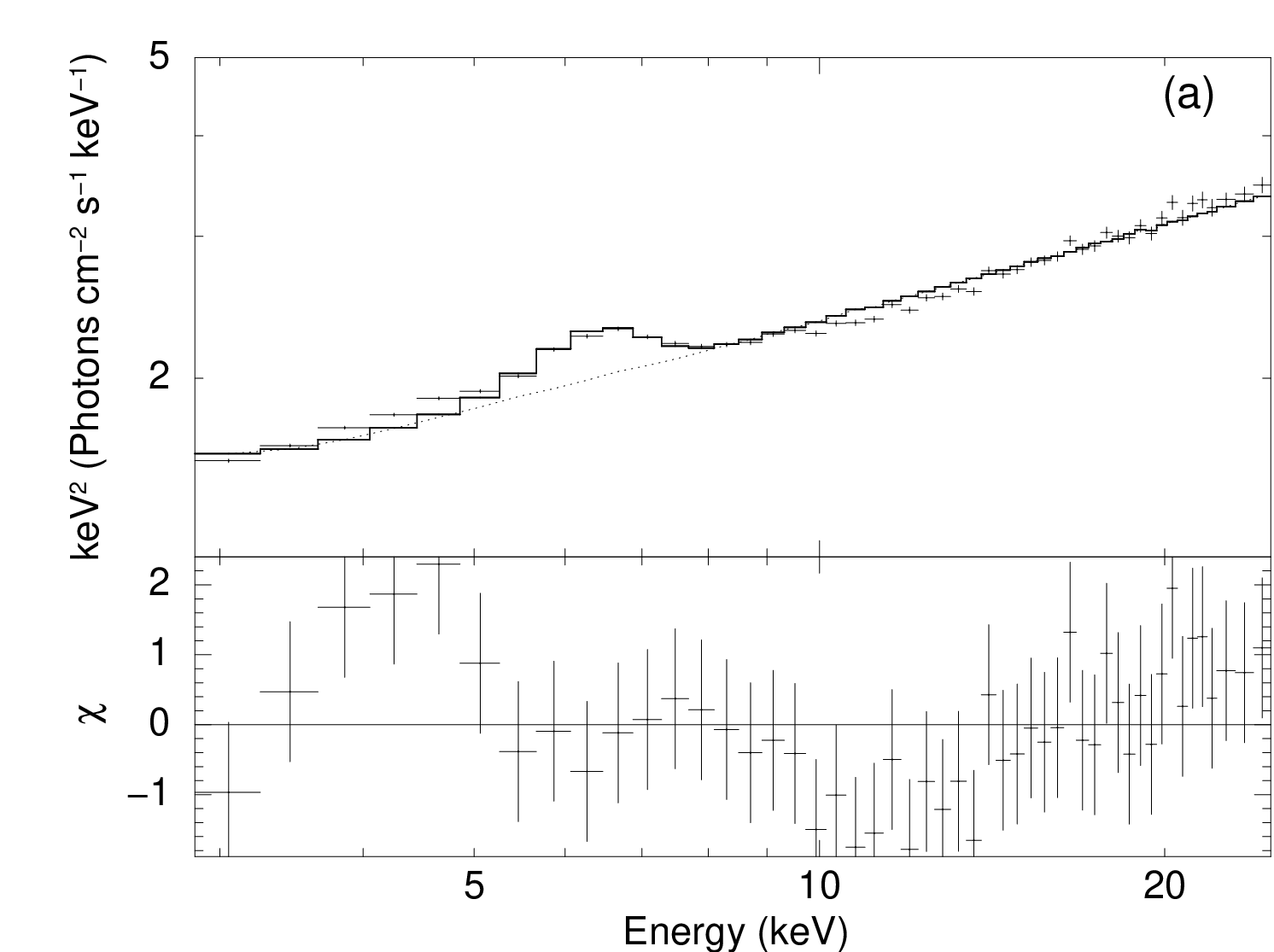}
\includegraphics[scale=0.6,angle=270,width=4.5truecm]{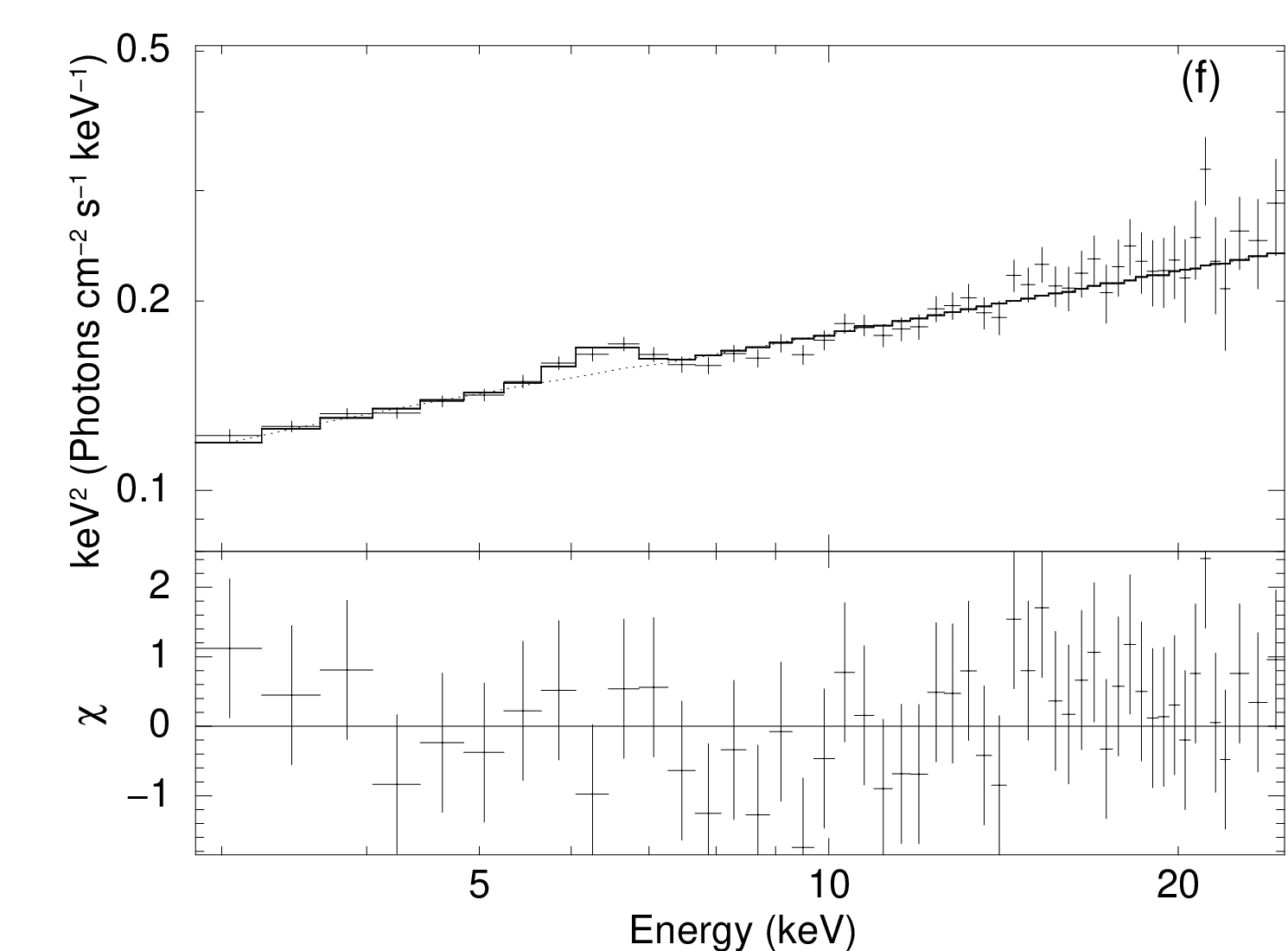}
}
\centerline{
\includegraphics[scale=0.6,angle=270,width=4.5truecm]{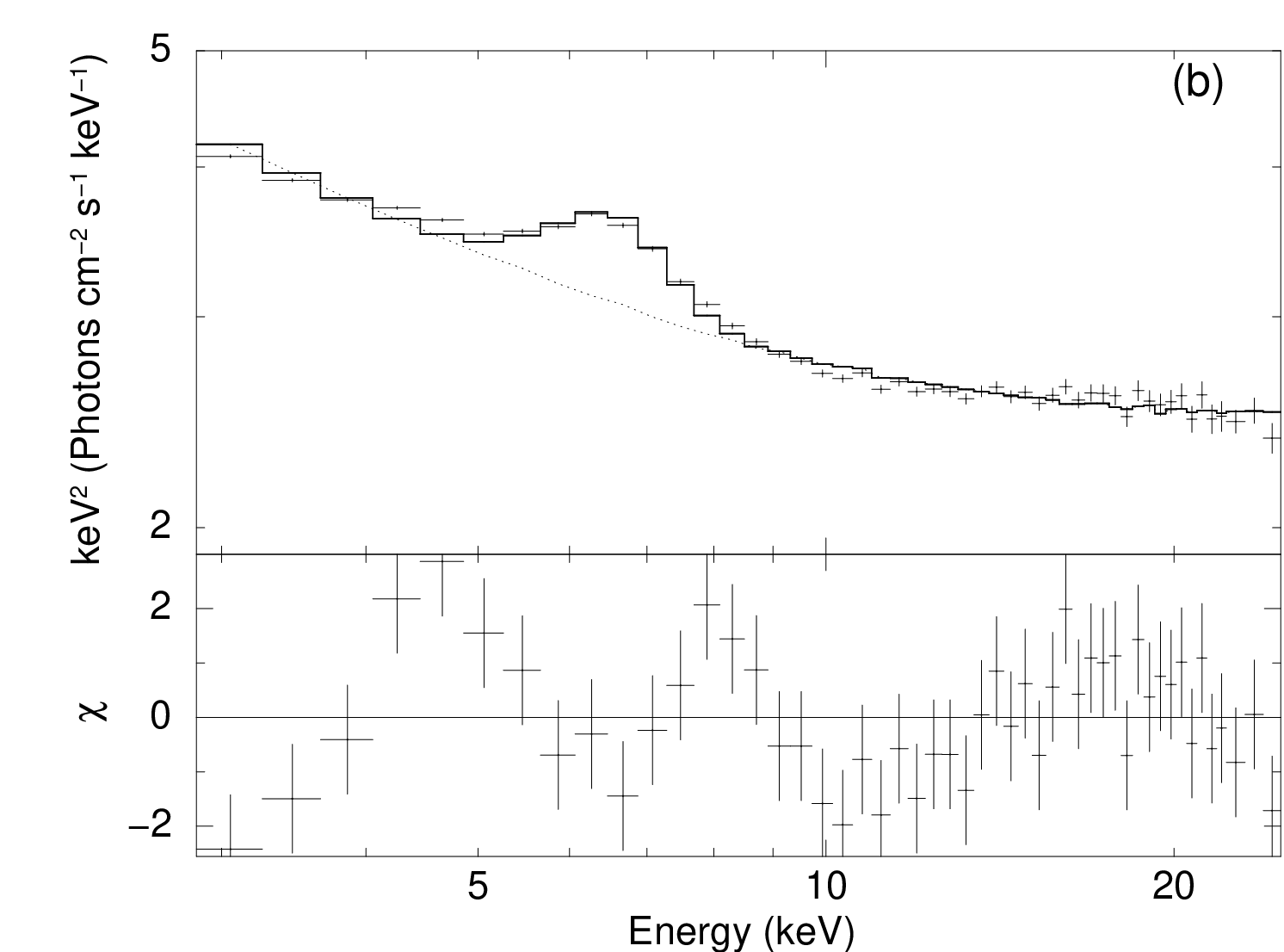}
\includegraphics[scale=0.6,angle=270,width=4.5truecm]{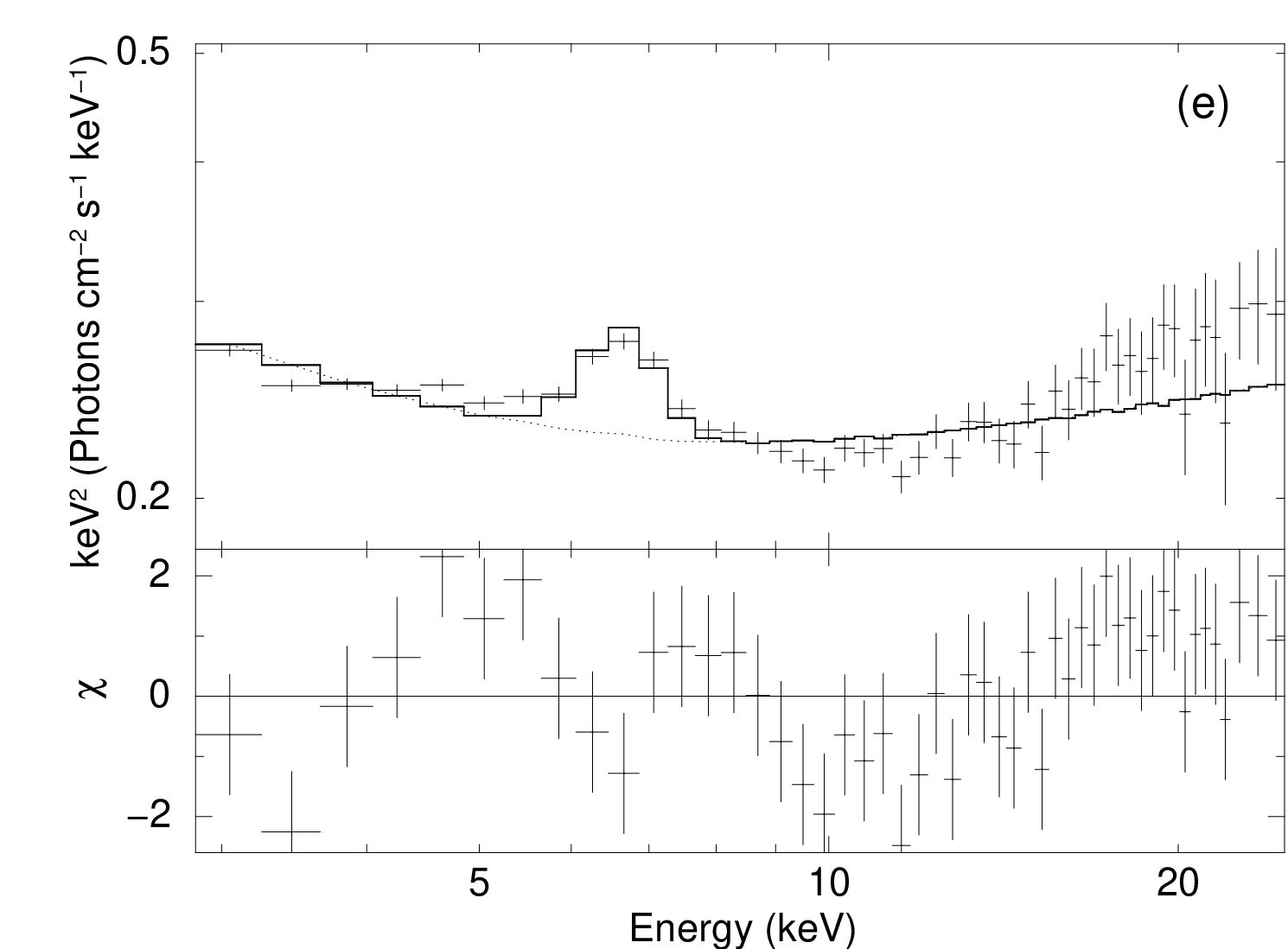}
}
\centerline{
\includegraphics[scale=0.6,angle=270,width=4.5truecm]{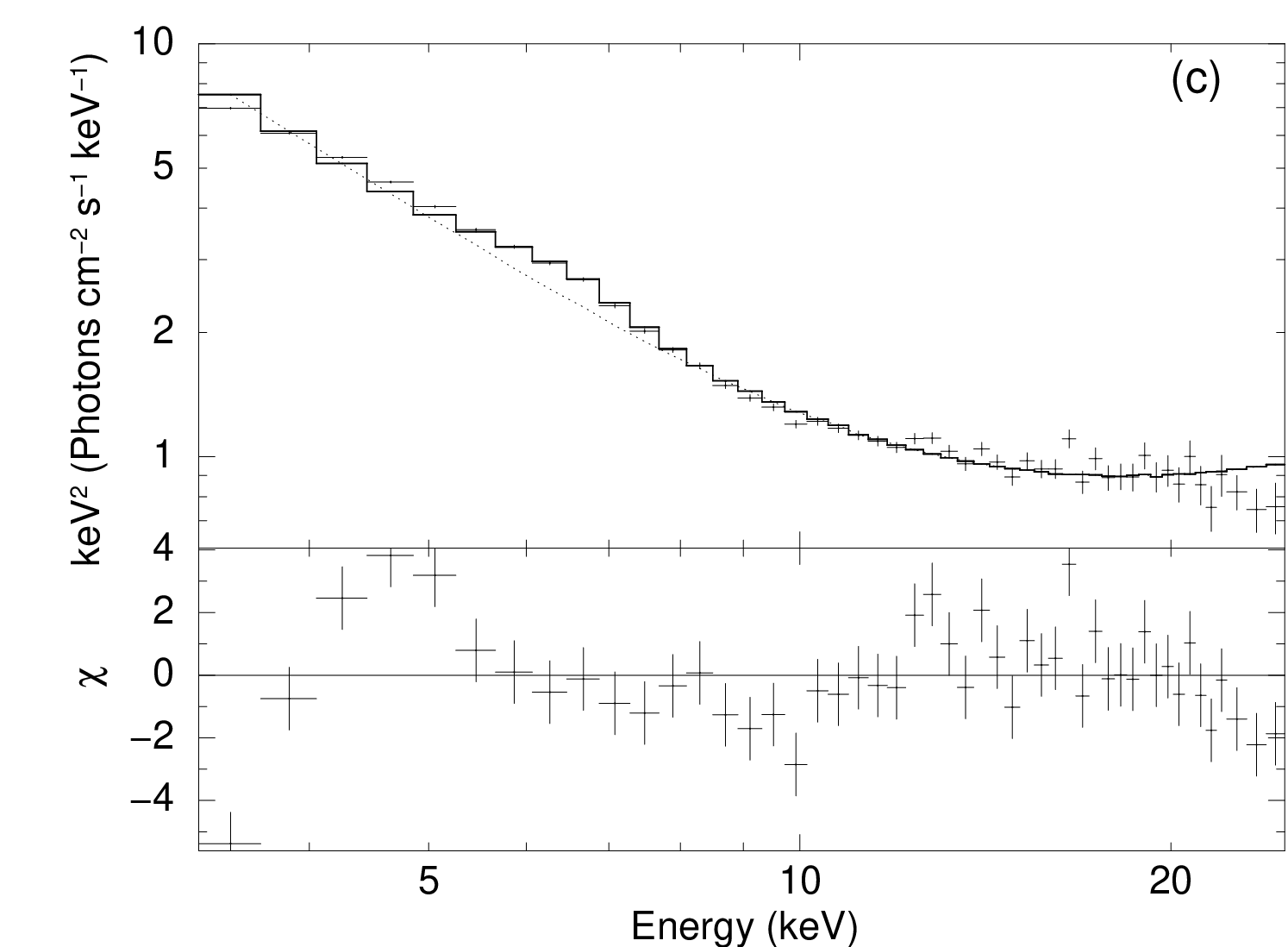}
\includegraphics[scale=0.6,angle=270,width=4.5truecm]{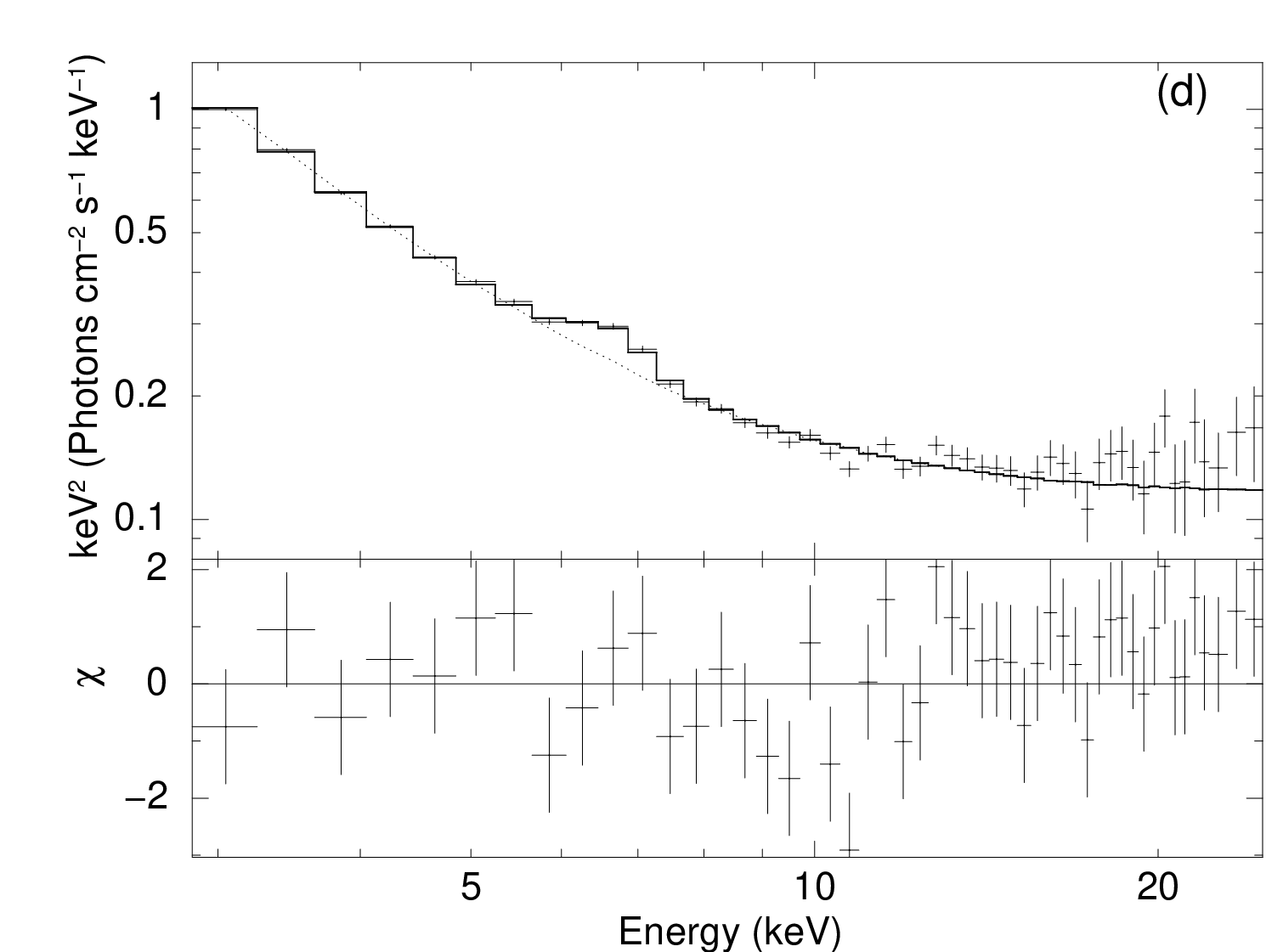}
}
\caption{TCAF model fitted $2.5-25$~keV PCA spectral flux, $E^2 f(E)$ 
in units of $keV^2~(Photons~cm^{-2}~s^{-1}~keV^{-1})$ 
with variation of $\Delta \chi$, selected from the six different spectral states whose results are marked with (a-f), 
presented in Table Appendix I. Left panel spectra are from rising phase and right panel spectra are from declining phase 
of the outburst. From top to bottom, hard, hard-intermediate and soft-intermediate states spectra are shown.
\label{fig4}}
\end{figure}

\noindent{\it (i) Hard State in Rising phase:}

RXTE PCA started monitoring the source nine days after reported X-ray flaring activity \citep{Tomsick10} 
observed with MAXI/GSC onboard ISS on 2010 January 03 \citep{Yamaoka10}. 
From the first observation day (2010 January 12; MJD = 55208), the source was in hard state with increasing 
contribution to total flux as well as percentage of non-thermal PL or sub-Keplerian halo rate (see, Figs. 1 \& 3). 
The source was observed in this spectral state till 2010 April 10 (MJD = 55296.49). In  this so-called hard 
state the flow is dominated by halo rates. QPOs are not observed for initial $\sim 68$ days up to 
2010 March 19 (MJD = 55274.35), which is quite unusual in transient 
black hole candidates \citep[see,]{D08,D13}. 
ARR increases monotonically and reached its maximum value ($\sim 7.38$) during rising phase (as well as the 
entire outburst) on 2010 April 11 (MJD = 55297.89), where we observe (Paper II) a transition from hard to 
hard-intermediate spectral state. ARR being maximum, the contribution from a Comptonized hot sub-Keplerian 
flow rate reached its maximum value with respect to thermally cool Keplerian flow rate. From Fig. 2, it is 
clear that during this phase of the outburst, the shock location ($X_s$) as well as shock height ($h_{shk}$) 
initially decrease with time (day) and then settle down to roughly a constant value. The shock is found to be 
weaker as the day progresses, initially slowly and then rapidly (see, Fig. 2b). Shock temperature ($T_{shk}$) 
initially increases, then settles down to a constant value, before decreasing rapidly, starting from the day 
when the first QPO is observed. While the shock location is generally decreasing, in the rising phase, it was 
not seen to be strictly monotonic as assumed to be the case in \citet{D10}. 
This may be because (a) the flow in CENBOL is not streamlined. Turbulence can cause shock movements a bit 
erratic. (b) Assumption of a constant accretion rate throughout the disk may not be correct in 
a rapidly evolving system. This may have forced the fitted shock location not to maintain monotonicity and
(c) Comptonization was assumed only due to the CENBOL region. Neglecting any contribution from jets
could easily change the fitted parameters either way. These causes affect the compression ratio of the 
shock. The shock location and compression ratio together decide the infall time scale in the post-shock
region which in turn decides the QPO frequency. 

\noindent{\it (ii) Hard-Intermediate State in the Rising phase :}

Source was in this state for next $\sim 6$ days after transition day (MJD = 55297.89). During this 
phase, QPO frequency is observed to increase monotonically as before from spectral transition day 
(MJD = 55297.89; where $1.241$~Hz QPO is observed) to 2010 April 17 (MJD = 55303.61; where maximum 
value of QPO frequency of $5.692$~Hz is observed). On the next day, on 2010 April 18 (MJD = 55304.72), 
observed QPO frequency ($5.739$~Hz) remained very similar and after that QPO appeared sporadically. 
This happens to be the day when the transition to the soft-intermediate state  took place (Paper II).
On this day, ARR reaches its lowest value compared to what it had for preceding 
more than three weeks. 
It is to be noted that during this spectral state, as the day progressed, supply of thermally 
cool Keplerian matter ($\dot{m_d}$) is increased, at the same time supply of sub-Keplerian flow ($\dot{m_h}$) 
is decreased. As a result of that, ARR is decreased monotonically. The overall spectra are still dominated by 
hard radiation from high halo rates (see, Figs. 1-3). 

\noindent{\it (iii) Soft-Intermediate State in the Rising phase:}

Source was observed in this spectral state for the next $\sim 26$ days from spectral transition day 
(2010 April 18; MJD = 55304.72), where sporadic QPOs of frequencies $\sim 4.7-6.6$~Hz are observed. 
During this spectral state, PCA count rates as well as total spectral flux/accretion rates are initially 
increased and then decreased (Paper II). This is because of initial increase in PL flux/sub-Keplerian rates 
and decrease in DBB flux/Keplerian rate. As a result, ARR varies within a short range $\sim 0.07 - 0.21$. 
In this state, the last prominent QPO of $4.687$~Hz is observed on 2010 May 14 (MJD = 55330.29). 
On the next day, 2010 May 15 (MJD = 55331.55) no QPO is observed, where both disk and halo rates 
are decreased suddenly from its previously observed values.
During this spectral state, flickering behaviors of shock temperature and compression ratio 
are observed (see, Fig. 2). As the disk rate rises, spectral index becomes very sensitive to the disk 
and halo parameters (CT95). It is possible that flickering of the fitted parameters is because of that.


\noindent{\it (iv) Soft State:}

Present version of fits file is not suitable to fit spectra from this spectral state as the flow essentially 
contains a single standard Keplerian disk component with a weak power-law component due to bulk motion 
Comptonization (which has not been included in this version of TCAF fits model) and non-thermal component 
\citep{Zdziarski01,CM06}. 
Still, we fitted three spectra including two spectral transition days with acceptable values of reduced 
$\chi^2$ ($<3$). The source stayed in this spectral state for a long period of time ($\sim 8$ months). 
On soft to soft-intermediate transition day, 2011 January 04 (MJD = 55565.83) ARR is observed to be $\sim 0.13$.

\noindent{\it (v) Soft-Intermediate State in Declining phase:}

The source was observed in this spectral state for about a month, where some variation in ARR
($0.13-0.16$) is observed, because of the reductions of $\dot{m_d}$ and $\dot{m_h}$ (see, Figs. 1-3). 
During this phase, similar to rising soft-intermediate spectral state, QPOs are observed to be 
sporadic (at $\sim 0.7-2$~Hz). On 2011 February 02 (MJD = 55594.90), another spectral transition 
from soft-intermediate to hard-intermediate state is observed, due to rise in $\dot{m_h}$ and decrease 
in $\dot{m_d}$, where a prominent QPO of frequency $6.420$~Hz is observed. 

\noindent{\it (vi) Hard-Intermediate State in Declining phase:}

Source was observed in this spectral state for the next $\sim 10$~days starting from spectral transition 
(soft-intermediate to hard-intermediate) day (2011 February 02; MJD = 55594.90). During this phase, similar 
to declining hard-intermediate state of other transient BHCs 
(e.g., \citealt{D08} for 2005 outburst of GRO~J1655-40; \citealt{D13} for 2010 \& 2011 outbursts of H~1743-322)
QPO frequencies are observed to decrease monotonically. As time passed, ARR increased monotonically due to 
daily rise in $\dot{m_h}$ rate and fall in $\dot{m_d}$ rate. ARR reaches its maximum value for the declining 
phase ($\sim 0.50$) on 2011 February 14 (MJD = 55606.90), where QPO of frequency $0.175$~Hz is observed. 
During this phase of the outburst, a slow increase in shock strength, shock height and 
temperature are observed. Shock also moved outward (see, Fig. 2). 

\noindent{\it (vii) Hard State in Declining phase:}

The source was observed in this spectral state till the end of observational data set (2011 March 06; MJD = 
55626.56). In this phase, prominent QPOs are observed only for two days, where QPO frequency 
is decreased from $0.175$~Hz to $0.136$~Hz. As the day progresses, both types of accretion rates 
($\dot{m_d}$ and $\dot{m_h}$) decrease due to less supply of fresh matter 
(reduced viscosity). Thus ARR decreases monotonically and reaches a value of
($\sim 0.22$) on the last day of the observation (MJD = 55626.56). During this state, a decrease in 
shock temperature with rapid rise in shock strength and height was seen. Also, 
shock is found to accelerate away from the black hole (see, Fig. 2). 

\section {Discussion and Concluding Remarks}

For the first time, spectral properties of GX~339-4 during its recent 2010-11 outburst have been studied 
in detail and the spectra were fitted using two component advective flow (TCAF) model after its inclusion 
as an additive table model in HEASARC's spectral analysis software package XSPEC. For the inclusion of the 
model in XSPEC, a {\it fits} file was generated with $\sim 4 \times 10^5$ model spectra created by varying 
input parameters in the CT95 code, modified to include general shock location and shock strengths. For data 
analysis, we use RXTE/PCA spectral data in $2.5-25$~keV energy band. We analyzed a total of $50$ observations 
spread over the entire outburst with TCAF model and compare our results with a combined DBB and PL model fitted 
results (Paper II). Variation of two component (Keplerian and sub-Keplerian) accretion rates extracted from 
TCAF model spectral fits are found to be consistent with DBB and PL model fluxes (see, Figs. 1 \& 3). The 
results are summarized in Table 'Appendix I'. What we found is that the ratio of sub-Keplerian halo rate and 
standard Keplerian disk rate, namely ARR, may play a very important role in deciding state transitions. 
For instance ARR was found to have a local sharp maximum, both in the rising and the declining phases, 
on the day of transition between the hard state and the hard-intermediate state. Hard and hard-intermediate 
states are dominated by the halo rate, whereas the other two states are dominated by the disk rate. Hard and 
intermediate (hard-intermediate, and soft-intermediate) spectra fit extremely well with our current TCAF 
{\it fits} file, since in these states, two components are prominent. Soft state is dominated by the standard 
Keplerian disk alone and better fits are achieved when $\dot{m_h} \rightarrow 0$. In future, the fits file will 
be expanded with finer grids for this regime. The fact that the present version of TCAF fits 
deteriorate in soft states is a direct indication of the absence of shocks in 
this state. Recently, \citet{MC13}, 
self-consistently solved transonic flow problem with Compton cooling. We hope to incorporate this into 
our code in future.

After fitting with TCAF, a clearer physical picture of what happens in an outburst emerges. Two types of 
accretion rates are found to vary independently. At the beginning of the outburst, spectra are dominated 
by low angular-momentum, thermally hot sub-Keplerian flow ($\dot{m_h}$) and the object is in a hard/low-hard 
state. As the outburst progresses, accretion rate of the Keplerian component ($\dot{m_d}$) increases and 
the object enters into a soft spectral state via two intermediate (hard-intermediate and soft-intermediate) 
spectral states. During the declining phase of the outburst, Keplerian disk recedes while the sub-Keplerian 
rate remains roughly the same or increases slightly, which makes the spectrum harder. So, at the end phase 
of the outburst, the source moves again in the hard state via two intermediate 
(soft-intermediate and hard-intermediate) spectral states. During this outburst 
of GX~339-4, maximum values of $\dot{m_h}$ (=8.14) and $\dot{m_d}$ (=5.99) 
are observed on 2010 April 11 (MJD = 55297.88) and 2010 April 18 (MJD = 55304.72) respectively. Since the 
sub-Keplerian matter moves roughly in a free-fall time scale, while the Keplerian disk  moves in a viscous time 
scale, this lag is expected. This delay may provide an estimate of viscosity in the Keplerian 
component. This will be dealt with elsewhere.

Evolution patterns of the location of the shock ($X_s$ in $r_g$ unit) and height of the shock 
($h_{shk}$ in $r_g$ unit) are more or less similar during this outburst of GX 339-4. During 
the rising phase of the outburst, as the day progresses, these two values generally decrease 
with time (day) and during the declining phase of the outburst, opposite scenario is observed. 
During rising phases of the outburst, the shock compression ratio ($R$) decreased initially 
slowly and then rapidly before it settled down to a lower value ($\sim 1.05$). During 
the declining phase, it increased almost monotonically. In the same way, CENBOL surface 
temperature ($T_{shk}$ in $10^{10}~K$) initially increased rapidly, then remained more or 
less constant at $\sim 4-5$. Interestingly, after the first observed QPO day (MJD = 55277), 
it rapidly decreased until the transition day of hard to hard-intermediate state. During declining hard state, 
$T_{shk}$ is observed to decrease monotonically, whereas during intermediate states $T_{shk}$ is found to be 
non-monotonic. In the present TCAF {\it fits} file, shock locations, shock strengths (inverse of the compression 
ratio), shock heights, and shock temperatures are all calculated simultaneously without taking care of outflows, 
if any. It is to be noted that in the present version of TCAF fits file, the input parameters
shock location and compression ratio both could be calculated from the fundamental flow equations, namely, from 
a set of energy and angular momentum of the accretion flow
\citep[C90a][]{MCD14}. 
In that case, our fit would yield more fundamental parameters. This will be done in 
next version of the TCAF model {\it fits} file. Variation of angular momentum would 
also give a measure of viscosity.
The general trend of the shock location is on an average the same as in Paper II. There is, however, some
discrepancy, because in Paper II, no spectral data was used to compute shock locations. Rather, an 
approximate algebraic relation between shock location and QPO frequency was used and that yielded quite high values 
of shock locations. The compression ratio was adjusted suitably to fit the QPO frequency variation.
In the present case, shock locations and compression ratios are obtained by TCAF fits and therefore are more realistic. 
We find that there are some fluctuations in shock locations on the top of monotonic behaviour. Part of the 
reason could be that along with Compton cooling which pushes the shock towards the black hole, 
presence and absence of winds (which is not considered here), which reduces density in 
CENBOL also affects the shock location. Another important assumption 
was that the accretion rate was treated as a constant throughout the disk \citep{DC10}, which is questionable in an evolving 
system. However, the general tendency of shock location remained acceptable and the behaviour is found to be 
consistent with anticipated behaviour. Another source of fluctuation could be that as the accretion rate goes up, 
the CENBOL switches from becoming an ion pressure supported to  a radiation pressure supported. So the size of the CENBOL
could grow at very soft state. In future, we will try to incorporate these aspects in our modified TCAF model {\it fits} file. 

Almost all the spectra are fitted with an Iron line of $\sim 6.2-6.7$~keV (by adding a Gaussian model 
component along with TCAF model), except those where $LW$ (line width in keV) and $LD$ (line depth) are absent
in Table I of Appendix. The non-Gaussian model fitted spectra are seen only in extreme two ends (hard states) 
of the outburst. Given that the resolution of RXTE/PCA is not good for line studies, a general nature of the Iron 
line evolution is observed. We do see a progressive widening of Iron line as we go from hard states to softer states. 
Similar variation is also observed for line depth. According to TCAF, lines are formed in the Keplerian disk just 
outside CENBOL and thus the solid angle to produce line widths would be large enough ($\sim 2\pi$). Also as the 
state becomes softer, rotational velocity increases as CENBOL moves in. This increases line broadening. 
Thermal broadening also increases, as the disk temperature rises. In future, we will include Iron line 
to generate {\it fits} file, since its width would be related to shock properties. 

Most natural assumption of the cause of an outburst, i.e., rise in flow rates, is possibly a rise in activity of 
the companion together with a rise in viscosity,
perhaps of magnetic origin, owing to enhanced magnetic activity in the companion, or convective turbulence
at the outer disk. If we assume that it is the 
sub-Keplerian component rate that is easy to change by the companion, the interpretations become easier. During 
the rising phase of an outburst, viscosity causes an increase in the accretion rate of Keplerian component 
since angular momentum of a sub-Keplerian flow could be easily re-distributed to produce a Keplerian disk
(Chakrabarti \& Molteni, 1995;
Giri \& Chakrabarti, 2013). Thus, a soft state is achieved in viscous time scale. As long as high 
(super-critical, C90ab) viscosity 
persists, the object stays in the soft state. When viscosity is reduced to sub-critical value, 
declining phase starts with a reduction of the Keplerian rate. Keplerian disk itself is 
retracted due to lack of conversion. A simple estimate of viscosity 
parameter suggests that $\alpha$ should increase from $0.001$ to $0.02$ from the first day 
to the day it came to the soft state \citep{MCD15}. 
These $\alpha$ values are consistent with what simulations with MRI indicate \citep{Arlt01,Masada09}. 
In the literature, a few models require higher viscosity to explain the origin of QPOs
and spectral states \citep{Mauche02,Titarchuk04}. 
This could be because their disk is Keplerian with high angular momentum, which requires
high viscosity ($\sim 0.1-0.5$) to transport angular momentum. On the contrary, TCAF assumes
a sub-Keplerian with low angular momentum, where lesser amount of viscosity
is sufficient to explain the spectral states and origin of QPO frequencies \citep{MCD15} through
shock formation and its oscillation. 

Generally type `C' QPOs are observed in hard, and hard-intermediate spectral states and type `B' QPOs 
are observed in soft-intermediate spectral states (see, van der Klis 2004). 
During hard and hard-intermediate spectral states, QPOs are observed continuously starting from the 
day when sudden rise in ARR is observed (see, Figs. 1 \& 3). It has been noticed that ARR reached 
its maximum values on the day the transition between hard to hard-intermediate states takes place. 
During soft-intermediate states, QPOs are observed sporadically as this state is slightly dominated 
by thermally cooler Keplerian flows, and QPOs are seen when sub-Keplerian halo rate rises
(see, Figs. 1 \& 3). This perception is more prominent in the rising soft-intermediate state 
because during the declining phase, total accretion rate is decreased as a whole. No LFQPOs are observed 
in soft states because oscillatory shock conditions which are required for the generation of QPOs 
are not satisfied in a `sub-sonic' Keplerian disk, which after all, dominates this state. As a result, 
accretion disk becomes thermally cool approximately at a constant temperature (see, Fig. 7a of Paper II). 
It appears that there is an association 
of a (though, sometimes weak) {\it peak} of ARR with a state 
transition. ARR has a local maximum on the day of transition between hard and hard-intermediate states.
This important finding was possible only after fitting with the TCAF solution. Of course, 
TCAF solution depends strongly on the behaviour of the companion. Unless the process of matter supply 
is well understood, predictability of the subject is limited to a duration much less than the stellar 
variability timescale.

According to the shock oscillation model by Chakrabarti and his collaborators in mid-90s, LFQPOs are 
originated due to oscillation of post-shock region \citep{MSC96,CAM04} 
when resonance occurs between infall time scale and cooling time scale in CENBOL or when Rankine-Hugoniot 
conditions are not satisfied to form a steady shock \citep{RCM97}. 
So, one can obtain the QPO frequency if one knows instantaneous shock location and compression ratio (assuming 
turbulence, which can increase infall time, is absent). 
Since we directly extract shock parameters from TCAF model fits, we should be able to predict frequency of low 
frequency QPOs, if present, and make a comparative study with POS model \citep{C05,C08,C09,D10,D13,N12}
fitted results. Indeed, we find that while going from hard-intermediate to soft-intermediate states, 
the rms value of QPOs is suddenly reduced (see, Table A.2 of Paper II). This points to the fact that the 
oscillating body (here the shock) is getting more and more diffused. Our finding of weakening of the 
shock strength corroborates with the general conclusion drawn from rms values. Indeed, weaker shocks could 
just be density enhancements due to centrifugal barrier and thus would produce 'B' type QPOs. 
This aspect will be looked into in a future work. 

\section*{Acknowledgments}

We are thankful to NASA/GSFC scientists and XTEhelp team members (specially to Dr. Keith A. Arnaud) 
for their kind help to write FORTRAN programs to generate model {\it fits} file by using theoretical 
(TCAF) model spectra, presented in this paper. This research has made use of RXTE data obtained through 
HEASARC Online Service, provided by the NASA/GSFC, in support of NASA High Energy Astrophysics Programs. 
We thank the anonymous referee of this paper for the very detailed comments which helped in
improving the quality of the paper. Also, S. Mondal acknowledges the support of CSIR-NET scholarship 
to carry forward his research works.


\clearpage


\begin{table}
\vskip 0.0cm
\addtolength{\tabcolsep}{-3.5pt}
\centering
\centerline{\large \bf Appendix I}
\vskip 0.1cm
\centerline {2.5-25 keV TCAF Model Fitted Parameters With QPOs}
\vskip 0.1cm
\begin{tabular}{lccccccccccccccc}
\hline
\hline
Obs&Id&MJD&$\dot{m_d}$&$\dot{m_h}$&ARR&$X_s$&R&$h_{shk}$&$T_{shk}$&$R_{z}$&$LW^\dagger$&$LD^\dagger$&QPO$^{\dagger\dagger}$&$\chi^2/DOF$\\
    &  & &($\dot{M}$$_{Edd}$)&($\dot{M}$$_{Edd}$)&&($r_g$)& &($r_g$)&($10^{10}K$)& (km)  &  (keV) &  & (Hz) & &   \\
 (1)&  (2)  & (3)  & (4)& (5) & (6) & (7) &  (8) & (9) & (10) & (11) & (12) & (13) & (14) & (15) \\
\hline

 1&X-01-00&55208.49&$0.281^{\pm0.010}$&$0.422^{\pm0.011}$&$1.508^{\pm0.093}$&$148.0^{\pm2.106}$&$1.670^{\pm0.011}$&$93.65^{\pm1.950}$&$0.885^{\pm0.012}$&$1.837^{\pm0.147}$&--  & --      & --  &40.99/42 \\
 2&X-03-00&55220.18&$0.382^{\pm0.013}$&$0.630^{\pm0.012}$&$1.649^{\pm0.088}$&$77.94^{\pm2.077}$&$1.629^{\pm0.018}$&$48.99^{\pm1.847}$&$1.668^{\pm0.041}$&$2.321^{\pm0.118}$&--  & --      & --  &72.31/45 \\
 3&X-04-00&55225.72&$0.493^{\pm0.014}$&$0.707^{\pm0.011}$&$1.434^{\pm0.063}$&$49.51^{\pm0.586}$&$1.644^{\pm0.029}$&$31.20^{\pm0.920}$&$2.659^{\pm0.063}$&$2.339^{\pm0.106}$&0.31& 1.09e-3 & --  &76.46/43 \\
 4&X-05-00&55232.27&$0.531^{\pm0.017}$&$0.732^{\pm0.018}$&$1.379^{\pm0.078}$&$32.31^{\pm0.463}$&$1.650^{\pm0.034}$&$20.38^{\pm0.712}$&$4.128^{\pm0.115}$&$2.093^{\pm0.108}$&0.44& 1.79e-3 & --  &51.11/42 \\
 5&X-07-00&55249.53&$0.817^{\pm0.036}$&$1.503^{\pm0.083}$&$1.839^{\pm0.183}$&$33.20^{\pm0.737}$&$1.635^{\pm0.021}$&$20.89^{\pm0.732}$&$3.993^{\pm0.096}$&$1.941^{\pm0.108}$&0.52& 2.01e-3 & --  &65.43/42 \\
 6&X-08-00&55257.71&$0.915^{\pm0.044}$&$1.542^{\pm0.139}$&$1.685^{\pm0.233}$&$24.32^{\pm1.081}$&$1.549^{\pm0.043}$&$15.02^{\pm1.085}$&$5.311^{\pm0.266}$&$1.779^{\pm0.119}$&0.56& 3.72e-3 & --  &41.86/42 \\
 7&X-09-00&55260.92&$0.937^{\pm0.073}$&$1.910^{\pm0.126}$&$2.038^{\pm0.293}$&$23.71^{\pm0.263}$&$1.354^{\pm0.027}$&$13.45^{\pm0.417}$&$4.602^{\pm0.117}$&$1.775^{\pm0.110}$&0.53& 4.59e-3 & --  &44.17/43 \\
 8&X-09-02&55264.26&$0.940^{\pm0.019}$&$2.174^{\pm0.182}$&$2.313^{\pm0.240}$&$21.43^{\pm0.848}$&$1.313^{\pm0.013}$&$11.79^{\pm0.583}$&$4.810^{\pm0.143}$&$1.775^{\pm0.127}$&0.58& 3.73e-3 & --  &48.48/43 \\
 9&X-10-05&55272.82&$1.102^{\pm0.037}$&$2.650^{\pm0.126}$&$2.405^{\pm0.195}$&$18.78^{\pm1.047}$&$1.270^{\pm0.012}$&$9.920^{\pm0.647}$&$5.096^{\pm0.190}$&$2.192^{\pm0.129}$&0.62& 7.28e-3 & --  &51.70/44 \\
10&X-11-00&55274.33&$0.888^{\pm0.026}$&$2.500^{\pm0.174}$&$2.815^{\pm0.278}$&$21.58^{\pm0.869}$&$1.292^{\pm0.020}$&$11.65^{\pm0.650}$&$4.601^{\pm0.164}$&$2.041^{\pm0.200}$&0.61& 6.27e-3 & --  &39.92/43 \\
11&X-11-02&55277.47&$0.928^{\pm0.011}$&$2.907^{\pm0.283}$&$3.133^{\pm0.342}$&$20.92^{\pm1.040}$&$1.321^{\pm0.017}$&$11.58^{\pm0.725}$&$4.999^{\pm0.189}$&$2.378^{\pm0.154}$&0.65& 1.08e-2 &0.102&43.55/43 \\
12$^a$&X-12-00&55281.59&$1.096^{\pm0.015}$&$3.303^{\pm0.157}$&$3.014^{\pm0.184}$&$22.35^{\pm0.497}$&$1.302^{\pm0.017}$&$12.18^{\pm0.430}$&$4.517^{\pm0.109}$&$3.088^{\pm0.123}$&0.69& 1.34e-2 &0.134&46.72/43 \\
13&X-13-00&55289.62&$1.107^{\pm0.050}$&$3.566^{\pm0.110}$&$3.221^{\pm0.245}$&$23.11^{\pm0.309}$&$1.106^{\pm0.033}$&$8.783^{\pm0.380}$&$2.122^{\pm0.077}$&$3.519^{\pm0.151}$&0.67& 1.82e-2 &0.261&72.14/43 \\
14&X-13-05&55292.78&$1.110^{\pm0.066}$&$5.542^{\pm0.339}$&$4.993^{\pm0.602}$&$22.36^{\pm0.692}$&$1.055^{\pm0.029}$&$6.417^{\pm0.375}$&$1.252^{\pm0.054}$&$4.735^{\pm0.135}$&0.70& 3.22e-2 &0.363&72.43/43 \\
15&X-14-01&55296.25&$1.262^{\pm0.013}$&$6.734^{\pm0.367}$&$5.336^{\pm0.346}$&$32.30^{\pm0.513}$&$1.068^{\pm0.042}$&$10.18^{\pm0.562}$&$1.031^{\pm0.049}$&$6.665^{\pm0.204}$&0.71& 1.89e-2 &1.026&64.21/43 \\
\hline
16&X-14-02&55297.88&$1.103^{\pm0.035}$&$8.143^{\pm0.438}$&$7.383^{\pm0.631}$&$19.28^{\pm0.125}$&$1.063^{\pm0.026}$&$5.877^{\pm0.182}$&$1.651^{\pm0.046}$&$5.454^{\pm0.185}$&0.77& 2.11e-2 &1.241&80.36/43 \\
17$^b$&X-14-06&55299.77&$1.921^{\pm0.084}$&$7.302^{\pm0.613}$&$3.801^{\pm0.485}$&$25.71^{\pm0.781}$&$1.171^{\pm0.015}$&$11.72^{\pm0.506}$&$2.732^{\pm0.076}$&$6.098^{\pm0.123}$&0.75& 2.65e-2 &2.420&71.49/43 \\
18&X-14-05&55301.79&$2.234^{\pm0.074}$&$5.936^{\pm0.754}$&$2.657^{\pm0.426}$&$29.69^{\pm0.662}$&$1.177^{\pm0.014}$&$13.70^{\pm0.469}$&$2.411^{\pm0.056}$&$11.62^{\pm0.122}$&0.76& 3.28e-2 &3.643&68.98/43 \\
19&X-15-00&55302.20&$2.632^{\pm0.111}$&$5.810^{\pm0.258}$&$2.207^{\pm0.191}$&$23.64^{\pm0.030}$&$1.186^{\pm0.018}$&$11.09^{\pm0.182}$&$3.162^{\pm0.050}$&$8.603^{\pm0.105}$&0.74& 2.81e-2 &4.177&70.68/43 \\
20&X-15-01&55303.61&$4.772^{\pm0.103}$&$1.259^{\pm0.036}$&$0.264^{\pm0.013}$&$46.40^{\pm0.310}$&$1.101^{\pm0.037}$&$17.29^{\pm0.697}$&$0.993^{\pm0.037}$&$12.26^{\pm0.164}$&0.68& 1.98e-2 &5.692&76.33/43 \\
\hline
21&X-15-02&55304.72&$5.999^{\pm0.377}$&$1.271^{\pm0.398}$&$0.212^{\pm0.080}$&$51.86^{\pm5.454}$&$1.073^{\pm0.054}$&$16.86^{\pm2.621}$&$0.675^{\pm0.069}$&$16.39^{\pm0.159}$&0.71& 2.56e-2 &5.739&112.7/43 \\
22&X-15-05&55307.03&$5.371^{\pm0.366}$&$0.955^{\pm0.230}$&$0.178^{\pm0.055}$&$52.92^{\pm6.125}$&$1.085^{\pm0.074}$&$18.36^{\pm3.377}$&$0.753^{\pm0.095}$&$21.83^{\pm0.122}$&0.78& 1.03e-2 & --  &105.3/43 \\
23&X-16-01&55310.34&$5.066^{\pm0.214}$&$0.628^{\pm0.088}$&$0.124^{\pm0.023}$&$65.84^{\pm7.246}$&$1.079^{\pm0.039}$&$22.14^{\pm3.237}$&$0.567^{\pm0.052}$&$53.70^{\pm0.119}$&0.80& 9.11e-3 & --  &121.6/42 \\
24$^c$&X-17-00&55316.12&$5.214^{\pm0.237}$&$0.924^{\pm0.104}$&$0.177^{\pm0.028}$&$52.59^{\pm6.788}$&$1.086^{\pm0.051}$&$18.33^{\pm3.227}$&$0.765^{\pm0.085}$&$21.02^{\pm0.139}$&0.83& 2.01e-2 &6.074&100.4/43 \\
25&X-17-03&55319.13&$4.808^{\pm0.203}$&$0.690^{\pm0.101}$&$0.144^{\pm0.027}$&$48.56^{\pm5.958}$&$1.067^{\pm0.049}$&$15.21^{\pm2.564}$&$0.670^{\pm0.072}$&$23.16^{\pm0.145}$&0.82& 2.18e-2 & --  &95.43/43 \\
26&X-17-05&55321.73&$4.922^{\pm0.110}$&$0.463^{\pm0.087}$&$0.094^{\pm0.020}$&$38.00^{\pm9.964}$&$1.170^{\pm0.044}$&$17.29^{\pm5.183}$&$1.817^{\pm0.306}$&$16.56^{\pm0.158}$&0.75& 1.90e-2 &5.250&102.8/43 \\
27&X-18-04&55327.04&$4.871^{\pm0.341}$&$0.413^{\pm0.076}$&$0.085^{\pm0.022}$&$33.13^{\pm8.474}$&$1.126^{\pm0.041}$&$13.48^{\pm3.940}$&$1.674^{\pm0.275}$&$14.04^{\pm0.153}$&0.85& 1.39e-2 &4.776&92.36/42 \\
28&X-19-00&55330.29&$4.945^{\pm0.275}$&$0.330^{\pm0.059}$&$0.067^{\pm0.016}$&$49.90^{\pm9.384}$&$1.196^{\pm0.051}$&$23.85^{\pm5.501}$&$1.517^{\pm0.207}$&$60.31^{\pm0.120}$&0.81& 7.66e-3 &4.687&98.30/42 \\
\hline
29&X-19-01&55331.55&$2.375^{\pm0.347}$&$0.191^{\pm0.061}$&$0.080^{\pm0.037}$&$34.30^{\pm6.916}$&$1.127^{\pm0.064}$&$14.00^{\pm3.619}$&$1.625^{\pm0.256}$&$24.86^{\pm0.155}$&0.95& 3.51e-3 & --  &95.24/43 \\
30&Y-01-01&55563.14&$2.209^{\pm0.085}$&$0.249^{\pm0.052}$&$0.113^{\pm0.028}$&$31.06^{\pm10.42}$&$1.172^{\pm0.024}$&$14.19^{\pm5.051}$&$2.255^{\pm0.424}$&$17.16^{\pm0.210}$&0.53& 2.01e-3 & --  &82.36/45 \\
\hline
31&Y-01-02&55565.83&$2.062^{\pm0.043}$&$0.262^{\pm0.017}$&$0.127^{\pm0.011}$&$29.71^{\pm6.988}$&$1.064^{\pm0.041}$&$9.120^{\pm2.496}$&$1.066^{\pm0.166}$&$11.48^{\pm0.303}$&0.46& 1.91e-3 & --  &54.05/42 \\
32&Y-02-02&55573.47&$2.143^{\pm0.110}$&$0.254^{\pm0.057}$&$0.119^{\pm0.033}$&$29.75^{\pm1.639}$&$1.120^{\pm0.019}$&$11.88^{\pm0.856}$&$1.801^{\pm0.080}$&$13.63^{\pm0.143}$&0.48& 2.77e-3 & --  &65.83/43 \\
33&Y-03-02&55580.62&$1.941^{\pm0.034}$&$0.274^{\pm0.022}$&$0.141^{\pm0.014}$&$29.60^{\pm1.554}$&$1.055^{\pm0.025}$&$8.495^{\pm0.647}$&$0.935^{\pm0.047}$&$9.251^{\pm0.161}$&0.51& 1.52e-3 & --  &53.78/43 \\
34$^d$&Y-04-02&55586.50&$1.809^{\pm0.010}$&$0.229^{\pm0.013}$&$0.127^{\pm0.008}$&$32.74^{\pm0.476}$&$1.061^{\pm0.022}$&$9.839^{\pm0.347}$&$0.924^{\pm0.026}$&$12.62^{\pm0.145}$&0.45& 1.54e-3 &2.158&32.74/43 \\
35&Y-05-00&55589.20&$1.697^{\pm0.014}$&$0.212^{\pm0.021}$&$0.125^{\pm0.013}$&$32.46^{\pm0.455}$&$1.061^{\pm0.027}$&$9.755^{\pm0.385}$&$0.932^{\pm0.030}$&$13.11^{\pm0.181}$&0.59& 1.32e-3 & --  &38.45/42 \\
36&Y-05-01&55591.62&$1.644^{\pm0.024}$&$0.222^{\pm0.029}$&$0.135^{\pm0.020}$&$29.59^{\pm0.174}$&$1.067^{\pm0.035}$&$9.267^{\pm0.358}$&$1.114^{\pm0.040}$&$10.74^{\pm0.183}$&0.63& 1.70e-3 &1.705&75.84/43 \\
37&Y-05-02&55593.51&$1.549^{\pm0.037}$&$0.253^{\pm0.006}$&$0.163^{\pm0.008}$&$29.18^{\pm0.472}$&$1.055^{\pm0.022}$&$8.374^{\pm0.310}$&$0.949^{\pm0.027}$&$7.548^{\pm0.202}$&0.60& 1.01e-3 &1.742&61.58/43 \\
\hline
38&Y-05-03&55594.90&$1.470^{\pm0.034}$&$0.289^{\pm0.010}$&$0.197^{\pm0.011}$&$23.34^{\pm0.368}$&$1.068^{\pm0.019}$&$7.357^{\pm0.247}$&$1.445^{\pm0.037}$&$2.668^{\pm0.172}$&0.58& 1.89e-3 &6.420&72.55/43 \\
39&Y-06-00&55597.26&$1.201^{\pm0.020}$&$0.390^{\pm0.017}$&$0.325^{\pm0.020}$&$27.22^{\pm0.449}$&$1.055^{\pm0.012}$&$7.812^{\pm0.218}$&$1.020^{\pm0.020}$&$3.244^{\pm0.208}$&0.52& 1.90e-3 &4.684&59.78/43 \\
40&Y-06-01&55598.67&$1.055^{\pm0.039}$&$0.344^{\pm0.017}$&$0.326^{\pm0.028}$&$29.78^{\pm0.354}$&$1.054^{\pm0.017}$&$8.476^{\pm0.237}$&$0.914^{\pm0.020}$&$5.334^{\pm0.104}$&0.43& 1.88e-3 &3.973&52.95/43 \\
41$^e$&Y-06-02&55601.89&$0.873^{\pm0.019}$&$0.325^{\pm0.014}$&$0.372^{\pm0.024}$&$26.01^{\pm0.240}$&$1.092^{\pm0.014}$&$9.327^{\pm0.206}$&$1.670^{\pm0.029}$&$3.193^{\pm0.112}$&0.46& 1.58e-3 &1.322&70.01/43 \\
42&Y-07-00&55603.99&$0.768^{\pm0.013}$&$0.318^{\pm0.022}$&$0.414^{\pm0.036}$&$37.53^{\pm1.007}$&$1.102^{\pm0.024}$&$14.04^{\pm0.683}$&$1.245^{\pm0.044}$&$2.825^{\pm0.133}$&0.42& 6.47e-4 &1.087&49.54/43 \\
43&Y-07-03&55604.90&$0.749^{\pm0.013}$&$0.324^{\pm0.011}$&$0.433^{\pm0.022}$&$37.91^{\pm0.877}$&$1.111^{\pm0.035}$&$14.68^{\pm0.802}$&$1.319^{\pm0.057}$&$2.440^{\pm0.115}$&0.41& 4.85e-4 &1.149&44.19/43 \\
\hline
44&Y-07-01&55606.90&$0.729^{\pm0.095}$&$0.365^{\pm0.039}$&$0.501^{\pm0.119}$&$30.49^{\pm0.387}$&$1.136^{\pm0.020}$&$12.78^{\pm0.387}$&$1.934^{\pm0.046}$&$1.676^{\pm0.108}$&0.32& 4.42e-4 &0.175&45.57/43 \\
45$^f$&Y-07-02&55607.76&$0.684^{\pm0.036}$&$0.340^{\pm0.015}$&$0.497^{\pm0.048}$&$42.45^{\pm1.229}$&$1.181^{\pm0.018}$&$19.74^{\pm0.873}$&$1.695^{\pm0.050}$&$1.366^{\pm0.118}$&0.39& 4.20e-4 &0.136&47.24/43 \\
46&Y-08-00&55611.61&$0.810^{\pm0.015}$&$0.257^{\pm0.021}$&$0.317^{\pm0.032}$&$49.67^{\pm4.026}$&$1.134^{\pm0.010}$&$20.69^{\pm1.860}$&$1.159^{\pm0.057}$&$1.971^{\pm0.138}$&0.20& 1.44e-4 & --  &42.92/43 \\
47&Y-08-01&55615.46&$0.739^{\pm0.078}$&$0.250^{\pm0.048}$&$0.338^{\pm0.101}$&$75.74^{\pm7.360}$&$1.311^{\pm0.048}$&$41.59^{\pm5.565}$&$1.311^{\pm0.112}$&$0.978^{\pm0.201}$& -- &  --     & --  &45.31/45 \\
48&Y-09-00&55617.55&$0.680^{\pm0.038}$&$0.181^{\pm0.025}$&$0.266^{\pm0.052}$&$86.00^{\pm8.680}$&$1.330^{\pm0.033}$&$47.95^{\pm6.030}$&$1.188^{\pm0.089}$&$1.168^{\pm0.148}$& -- &  --     & --  &35.24/45 \\
49&Y-09-02&55622.48&$0.628^{\pm0.038}$&$0.149^{\pm0.021}$&$0.237^{\pm0.048}$&$107.5^{\pm10.21}$&$1.399^{\pm0.040}$&$62.66^{\pm7.743}$&$1.036^{\pm0.079}$&$1.140^{\pm0.203}$& -- &  --     & --  &29.40/45 \\
50&Y-10-01&55626.55&$0.609^{\pm0.048}$&$0.136^{\pm0.089}$&$0.223^{\pm0.164}$&$133.6^{\pm14.92}$&$1.413^{\pm0.048}$&$78.45^{\pm11.43}$&$0.844^{\pm0.076}$&$1.175^{\pm0.218}$& -- &  --     & --  &27.61/45 \\
\hline
\end{tabular}
\noindent{
\leftline {Here X=95409-01, Y=96409-01 mean the initial part of the observation Ids, and (a-f) mark TCAF model fitted results for six different states, presented in Fig. 4.}
\leftline {Intermediate horizontal lines mark state transitions from HS$\rightarrow$HIMS, HIMS$\rightarrow$SIMS, SIMS$\rightarrow$SS, SS$\rightarrow$SIMS, SIMS$\rightarrow$HIMS, and HIMS$\rightarrow$HS respectively.}
\leftline {$\dot{m_h}$, and $\dot{m_d}$ represent TCAF model fitted sub-Keplerian (halo) and Keplerian (disk) rates in Eddington rate respectively. $X_s$ (in Schwarzschild radius $r_g$), and}
\leftline {$R$ are the model fitted shock location and compression ratio values respectively. $h_{shk}$ (in $r_g$) and $T_{shk}$ (in $10^{10}~K$) are the shock height and temperature values}
\leftline {derived from Eqs. 4 \& 5  respectively. $R_z$ (in km) represents effective height of the Keplerian component at the pre-shock region.}
\leftline {$^\dagger$ LW and LD represent Gaussian model fitted Iron line ($\sim 6.2 - 6.7$ keV) width and depth respectively.}
\leftline {$^{\dagger\dagger}$ Here, frequencies of the principal QPO in Hz are presented. DOF means degrees of freedom of the model fit.}
}
\end{table}	


\begin{thebibliography}{}
\bibitem[\protect\citeauthoryear{Allured et al.}{2013}]{Allured13}Allured, R., Tomsick, J. A., Kaaret, P. \& Yamaoka, K., 2013, ApJ, 774, 135
\bibitem[\protect\citeauthoryear{Arlt \& Rüdiger}{2001}]{Arlt01}Arlt, R. \& Rüdiger, G., 2001, A \& A, 374, 1035
\bibitem[\protect\citeauthoryear{Belloni et al.}{2005}]{Belloni05} Belloni, T., Homan, J., Casella, P., et al., 2005, A\&A, 440, 207
\bibitem[\protect\citeauthoryear{Buxton et al.}{2012}]{Buxton12} Buxton, M. M., Bailyn, C. D., \& Capelo, H. L., et al., 2012, AJ, 143, 130
\bibitem[\protect\citeauthoryear{Cadolle et al.}{2012}]{Cadolle12} Cadolle, B. M. et al., 2012, A \& A, 544, 2 
\bibitem[\protect\citeauthoryear{Chakrabarti}{1989}]{C89} Chakrabarti, S. K., 1989, MNRAS, 340, 7 (C89)
\bibitem[\protect\citeauthoryear{Chakrabarti}{1990a}]{C90a} Chakrabarti, S. K., 1990a, ``Theory of Transonic Astrophysical Flows", World Scientific (Singapore) (C90a)  
\bibitem[\protect\citeauthoryear{Chakrabarti}{1990b}]{C90b} Chakrabarti, S. K., 1990b, ApJ, 362, 406 (C90b)
\bibitem[\protect\citeauthoryear{Chakrabarti \& Molteni}{1995}]{CM95} Chakrabarti, S. K. \& Molteni, D., 1995, MNRAS, 272, 80 (CM95)
\bibitem[\protect\citeauthoryear{Chakrabarti \& Titarchuk}{1995}]{CT95} Chakrabarti, S. K. \& Titarchuk, L. G., 1995, ApJ, 455, 623 (CT95)
\bibitem[\protect\citeauthoryear{Chakrabarti}{1996}]{C96} Chakrabarti, S. K., 1996, ApJ, 464, 664
\bibitem[\protect\citeauthoryear{Chakrabarti}{1997}]{C97}Chakrabarti, S. K., 1997, ApJ, 484, 313 (C97)
\bibitem[\protect\citeauthoryear{Chakrabarti}{1999}]{C99} Chakrabarti, S.K., 1999, A \& A, 351, 185
\bibitem[\protect\citeauthoryear{Chakrabarti, Acharyya \& Molteni}{2004}]{CAM04} Chakrabarti, S. K., Acharyya, K., \& Molteni, D., 2004, A\&A, 421, 1
\bibitem[\protect\citeauthoryear{Chakrabarti et al.}{2005}]{C05} Chakrabarti, S. K., Nandi, A., \& Debnath, D., et al., 2005, IJP 79(8), 841 (astro-ph/0508024)
\bibitem[\protect\citeauthoryear{Chakrabarti \& Mandal}{2006}]{CM06}Chakrabarti, S. K. \& Mandal, S. 2006, ApJ, 642, L49
\bibitem[\protect\citeauthoryear{Chakrabarti et al.}{2008}]{C08} Chakrabarti, S. K., Debnath, D., \& Nandi, A., et al., 2008, A\&A, 489, L41 
\bibitem[\protect\citeauthoryear{Chakrabarti et al.}{2009}]{C09} Chakrabarti, S. K., Dutta, B. G. \& Pal, P. S., 2009, MNRAS, 394, 1463
\bibitem[\protect\citeauthoryear{Corbel et al.}{2013a}]{Corbel13a} Corbel, S., Coriat, M., \& Brocksopp, C., et al., 2013a, MNRAS, 428, 2500
\bibitem[\protect\citeauthoryear{Corbel et al.}{2013b}]{Corbel13b} Corbel, S., Aussel, H., Broderick, J. W., et al., 2013b, MNRAS, 431, 107
\bibitem[\protect\citeauthoryear{Debnath et al.}{2008}]{D08} Debnath, D., Chakrabarti, S. K., Nandi, A., et al., 2008, BASI, 36, 151
\bibitem[\protect\citeauthoryear{Debnath et al.}{2010}]{D10} Debnath, D., Chakrabarti, S. K., \& Nandi, A., 2010, A\&A, 520, 98 (Paper I)
\bibitem[\protect\citeauthoryear{Debnath et al.}{2013}]{D13} Debnath, D., Chakrabarti, S. K., \& Nandi, A., 2013, AdSpR, 52, 2143
\bibitem[\protect\citeauthoryear{Debnath, Chakrabarti \& Mondal}{2014}]{DCM14} Debnath, D., Chakrabarti, S. K., \& Mondal, S., 2014, MNRAS, 440, L121 (DCM14)
\bibitem[\protect\citeauthoryear{Dincer et al.}{2012}]{Dincer12} Dincer, T., Kalemci, E., \& Buxton, M. M., et al., 2012, ApJ, 753, 55
\bibitem[\protect\citeauthoryear{Dutta \& Chakrabarti}{2010}]{DC10} Dutta, B. G., \& Chakrabarti, S. K., 2010, MNRAS, 404, 2136
\bibitem[\protect\citeauthoryear{Ebisawa, Titarchuk \& Chakrabarti}{1996}]{ETC96} Ebisawa, K., Titarchuk, L., \& Chakrabarti, S. K., 1996, PASJ, 48, 59
\bibitem[\protect\citeauthoryear{Giri \& Chakrabarti}{2013}]{GC13} Giri, K., \& Chakrabarti, S. K., 2013, MNRAS, 430, 2836
\bibitem[\protect\citeauthoryear{Giri, Garain \& Chakrabarti}{2014}]{GGC14} Giri, K., \& Chakrabarti, S. K., 2014, MNRAS (submitted)
\bibitem[\protect\citeauthoryear{Haardt \& Maraschi}{1993}]{Haardt93} Haardt, F., \& Maraschi, L., 1993, ApJ, 413, 507
\bibitem[\protect\citeauthoryear{Homan \& Belloni}{2005}]{Homan05} Homan, J. \& Belloni, T., 2005, Ap\&SS, 300, 107
\bibitem[\protect\citeauthoryear{Hua \& Titarchuk}{1995}]{Hua95} Hua, X. \& Titarchuk, L.G. 1995, ApJ, 449, 188
\bibitem[\protect\citeauthoryear{Hynes et al.}{2003}]{Hynes03} Hynes, R. I., Steeghs, D., \& Casares, J., et al., 2003, ApJ, 583, L95
\bibitem[\protect\citeauthoryear{Hynes et al.}{2004}]{Hynes04} Hynes, R. I., Steeghs, D., \& Casares, J., et al., 2004, ApJ, 609, 317
\bibitem[\protect\citeauthoryear{Landau \& Lifshitz}{1959}]{Landau59} Landau, L. D. \& Lifshitz, E. M., 1959, ``Fluid Mechanics", Oxford
\bibitem[\protect\citeauthoryear{Maccarone \& Coppi}{2003}]{Maccarone03} Maccarone, T. J. \& Coppi, P. S., 2003, MNRAS, 338, 189
\bibitem[\protect\citeauthoryear{Mandal \& Chakrabarti}{2010}]{MC10} Mandal, S. \& Chakrabarti, S. K., 2010, ApJ, 710, L147
\bibitem[\protect\citeauthoryear{Markert et al.}{1973}]{Markert73} Markert, T. H., Canizares, C. R., \& Clark, G. W., et al., 1973, ApJ, 184, L67
\bibitem[\protect\citeauthoryear{Masada \& Sano}{2009}]{Masada09}Masada, Y. \& Sano, T. 2009,  IAU Symposium No.  259, 121
\bibitem[\protect\citeauthoryear{Mauche}{2002}]{Mauche02}Mauche, C., 2002, ApJ, 580, 423
\bibitem[\protect\citeauthoryear{McClintock \& Remillard}{2006}]{MR06} McClintock, J. E., \& Remillard, R. A., 2006, in Compact Stellar X-ray Sources, Cambridge, Astrophysical Ser., vol. 39, ed. W. Lewin \& M. van der Klis (Cambridge Univ. Press), 157
\bibitem[\protect\citeauthoryear{Mondal \& Chakrabarti}{2013}]{MC13} Mondal, S., \& Chakrabarti, S. K., 2013, MNRAS, 431, 2716
\bibitem[\protect\citeauthoryear{Mondal, Debnath \& Chakrabarti}{2014a}]{MDC14} Mondal, S., Debnath, D., \& Chakrabarti, S. K., 2014a, ApJ, 786, 4 (MDC14)
\bibitem[\protect\citeauthoryear{Mondal, Chakrabarti \& Debnath}{2014b}]{MCD14} Mondal, S., Chakrabarti, S. K., \& Debnath, D., 2014b, Ap\&SS, 353, 223
\bibitem[\protect\citeauthoryear{Mondal, Chakrabarti \& Debnath}{2015}]{MCD15} Mondal, S., Chakrabarti, S. K., \& Debnath, D., 2015, ApJ, 798, 57
\bibitem[\protect\citeauthoryear{Molteni et al.}{1994}]{Molteni94} Molteni, D., Lanzafame, G. \& Chakrabarti, S.K., 1994, ApJ, 425, 161
\bibitem[\protect\citeauthoryear{Molteni, Sponholz \& Chakrabarti}{1996}]{MSC96} Molteni, D., Sponholz, H. \& Chakrabarti, S. K., 1996, ApJ, 457, 805
\bibitem[\protect\citeauthoryear{Motta et al.}{2009}]{Motta09} Motta, S., Belloni, T., \& Homan, J., 2009, MNRAS, 400, 1603
\bibitem[\protect\citeauthoryear{Motta et al.}{2011}]{Motta11} Motta, S., Munoz-Darias, T., \& Casella, P., et al., 2011, MNRAS, 418, 2292
\bibitem[\protect\citeauthoryear{Nandi et al.}{2012}]{N12} Nandi, A., Debnath, D., \& Mandal, S., et al., 2012, A\&A, 542, 56 (Paper II)
\bibitem[\protect\citeauthoryear{Nowak et al.}{1999}]{Nowak99} Nowak, M. A., Wilms, J., \& Dove, J. B., 1999, ApJ, 517, 355
\bibitem[\protect\citeauthoryear{Okuda et al.}{2007}]{Okuda07} Okuda, T., Teresi, V. \& Molteni, D., 2007, MNRAS 377, 1431	
\bibitem[\protect\citeauthoryear{Paczy\'nski \& Witta}{1980}]{PW80} Paczy\'nski, B., \& Witta, P. J., 1980, A\&A, 88, 23
\bibitem[\protect\citeauthoryear{Paczy\'nski \& Bisnovatyi-Kogan}{1981}]{PB81} Paczy\'nski, B., \& Bisnovatyi-Kogan, G., 1981, Acta Astron., 31, 283
\bibitem[\protect\citeauthoryear{Rahoui et al.}{2012}]{Rahoui12} Rahoui, F., Coriat, M., \& Corbel, S., et al., 2012, MNRAS, 422, 2202
\bibitem[\protect\citeauthoryear{Remillard \& McClintock}{2006}]{RM06} Remillard, R. A., \& McClintock, J. E., 2006, ARA\&A, 44, 49
\bibitem[\protect\citeauthoryear{Ryu, Chakrabarti, \& Molteni}{1997}]{RCM97} Ryu, D., Chakrabarti, S. K. \& Molteni, D., 1997, ApJ, 474, 378
\bibitem[\protect\citeauthoryear{Shakura \& Sunyaev}{1973}]{SS73} Shakura, N. I., \& Sunyaev, R. A., 1973, A\&A, 24, 337 (SS73)
\bibitem[\protect\citeauthoryear{Shidatsu et al.}{2011a}]{Shidatsu11a} Shidatsu, M., Ueda, Y., Tazaki, F., et al., 2011, PASJ, 63, 785
\bibitem[\protect\citeauthoryear{Shidatsu et al.}{2011b}]{Shidatsu11b} Shidatsu, M., Ueda, Y., Nakahira, S., et al., PASJ, 63, 803
\bibitem[\protect\citeauthoryear{Shimura \& Takahara}{1995}]{ST95} Shimura, T., \& Takahara, F., 1995, ApJ, 445, 780
\bibitem[\protect\citeauthoryear{Singh \& Chakrabarti}{2011}]{SC11} Singh, C. B., \& Chakrabarti, S. K., 2011, MNRAS, 410, 2414
\bibitem[\protect\citeauthoryear{Sobolev}{1975}]{Sobolev75}Sobolev, V.V. 1975, Light Scattering in Planetary Atmosphere (Oxford: Pergamon)
\bibitem[\protect\citeauthoryear{Stiele et al.}{2011}]{Stiele11} Stiele, H., Motta, S., \& Mu\~{n}oz-Darias, T., et al., 2011, MNRAS, 418, 1746 
\bibitem[\protect\citeauthoryear{Sunyaev \& Titarchuk}{1980}]{ST80} Sunyaev, R.A., \& Titarchuk, L. G., 1980, ApJ, 86, 121
\bibitem[\protect\citeauthoryear{Sunyaev \& Titarchuk}{1985}]{ST85} Sunyaev, R.A., \& Titarchuk, L. G., 1985, A\&A, 143, 374
\bibitem[\protect\citeauthoryear{Tomsick}{2010}]{Tomsick10} Tomsick, J. A., 2010, ATel, 2384, 1
\bibitem[\protect\citeauthoryear{Titarchuk \& Fiorito}{2004}]{Titarchuk04}Titarchuk, L \& Fiorito, R., 2004, ApJ, 612, 988
\bibitem[\protect\citeauthoryear{Titarchuk \& Seiﬁna}{2009}]{Titarchuk09}Titarchuk, L.G., \& Seiﬁna, E. 2009, ApJ, 706, 1463
\bibitem[\protect\citeauthoryear{Titarchuk \& Zannias}{1998}]{Titarchuk98}Titarchuk, L.G., \& Zannias, T. 1998, ApJ, 493, 863
\bibitem[\protect\citeauthoryear{van der Klis}{2004}]{vdK04} van der Klis, M., 2004, AN, 326, 798
\bibitem[\protect\citeauthoryear{Yamaoka et al.}{2010}]{Yamaoka10} Yamaoka, K., Nakahira, S., \& Mihara, T., et al., 2010, ATel, 2380, 1
\bibitem[\protect\citeauthoryear{Yan \& Yu}{2012}]{Yan12} Yan, Z. \& Yu, W., 2012, MNRAS, 427, L11
\bibitem[\protect\citeauthoryear{Zdziarski et al.}{2001}]{Zdziarski01}Zdziarski, A. A., Grove, J. E., Poutanen, J., Rao, A. R., \& Vadawale, S. V.,  2001, ApJ, 554, L45
\bibitem[\protect\citeauthoryear{Zdziarski et al.}{2003}]{Zdziarski03} Zdziarski, A. A., Lubinski, P., \& Gilfanov, M., et al., 2003, MNRAS, 342, 355

\end{thebibliography}
\end{document}